\documentclass[twocolumn]{aastex631}  
\usepackage{graphicx}
\usepackage{txfonts}

\usepackage{natbib}
\begin{document}

\title{Dayside Clouds and an Elevated C/O Ratio in the Atmosphere of the Ultra-hot Jupiter WASP-19b}

\author[0000-0001-8018-0264]{Suman Saha}
\affiliation{Instituto de Estudios Astrofísicos, Facultad de Ingeniería Ciencias, Universidad Diego Portales, Av. Ejército Libertador 441, Santiago, Chile}
\affiliation{Centro de Excelencia en Astrofísica y Tecnologías Afines (CATA), Camino El Observatorio 1515, Las Condes, Santiago, Chile}
	
\correspondingauthor{Suman Saha}
\email{suman.saha@mail.udp.cl}

\author[0000-0003-2733-8725]{James S. Jenkins}
\affiliation{Instituto de Estudios Astrofísicos, Facultad de Ingeniería Ciencias, Universidad Diego Portales, Av. Ejército Libertador 441, Santiago, Chile}
\affiliation{Centro de Excelencia en Astrofísica y Tecnologías Afines (CATA), Camino El Observatorio 1515, Las Condes, Santiago, Chile}

\accepted{for publication in The Astronomical Journal}

\begin{abstract}
Ultra-hot Jupiters (UHJs) offer exceptional opportunities for detailed atmospheric characterization via emission spectroscopy. We present a comprehensive analysis of the dayside atmosphere of WASP-19b—one of the shortest-period UHJs—using archival JWST NIRSpec/PRISM observations, leveraging its broad panchromatic wavelength coverage (0.6-5.2 $\mu$m). Using atmospheric retrievals, we report robust detections of H$_2$O (20.41 $\sigma$) and CO (4.79 $\sigma$), along with marginal evidence for CO$_2$ (2.11 $\sigma$) and VO (2.81 $\sigma$). Our retrievals also place strong constraints on the abundances of SiO and TiO, although their presence is not statistically confirmed. Furthermore, our analysis reveals strong evidence for clouds in this ultra-hot dayside atmosphere (4.3 $\sigma$), with indications of silicon dioxide (silica/quartz; SiO$_2$(s)) cloud formation—making WASP-19b the first UHJ with a statistically significant cloud detection. Leveraging the well-constrained molecular abundances, we infer a dayside C/O ratio of 0.77 $\pm$ 0.16, a potentially super-solar value consistent with emerging trends among UHJs and suggestive of possible oxygen sequestration through cloud condensation. Our findings place WASP-19b as a key benchmark for modeling dayside atmospheric processes and evolutionary dynamics in extremely irradiated exoplanets.
\end{abstract}

\section{Introduction}

Emission spectroscopy of transiting exoplanets is rapidly emerging as one of the most powerful techniques for probing the physical structures and chemical compositions of their atmospheres. Ultra-hot Jupiters (UHJs)—gas giants with dayside temperatures exceeding 2000K—are particularly well-suited for such studies due to their strong thermal emission and inflated atmospheres \citep[e.g.,][]{2018AJ....156...10M, 2023Natur.620..292C, 2023AJ....165..211R, saha2025a}. Near-to mid-infrared emission spectra of these extreme worlds can reveal the precise atmospheric abundances of key molecular species—such as H$_2$O, CO, CO$_2$, and SiO—as well as place constraints on cloud composition and vertical structure \citep[e.g.,][]{2024ApJ...973L..41I, 2025NatAs.tmp...56C, 2025A&A...700A..45S, saha2025a}. Since UHJs may have formed through unique evolutionary pathways, such atmospheric characterization also offers valuable insights into their origins and dynamical histories \citep[e.g.,][]{2016ApJ...832...41M, 2017AJ....153...83B}.

WASP-19b \citep{2010ApJ...708..224H}, which was the shortest-period transiting exoplanet known at the time of its discovery, orbits its G8V-type host star in just $\sim$0.79 days. With an equilibrium temperature of 2113K and a radius of 1.41 R$_J$ \citep{2020A&A...636A..98C}, it lies well within the UHJ population, orbiting the coolest host star among them all. These characteristics make WASP-19b a particularly compelling target for atmospheric follow-up studies. Several previous attempts have been made to characterize its atmosphere using both ground- and space-based observations, despite the challenge posed by its relatively faint host star (V$_{mag}$ $\approx$ 12.25).

Previous studies using HST and Spitzer have reported the detection of water vapor in the atmosphere of WASP-19b, along with the presence of hazes absorbing at shorter wavelengths \citep{2024AJ....168..296T, 2023ApJS..269...31E, 2022ApJS..260....3C, 2016ApJ...823..122W, 2013MNRAS.434.3252H, 2013ApJ...779..128M, 2013MNRAS.430.3422A}. Ground-based optical to near-infrared observations using VLT/FORS2, Magellan/IMACS, and VLT/ESPRESSO have also suggested the possible presence of TiO and reaffirmed the existence of high-altitude clouds and hazes \citep{2017Natur.549..238S, 2019MNRAS.482.2065E, 2021MNRAS.505..435S}. However, these earlier studies were unable to robustly constrain key chemical properties such as the metallicity or carbon-to-oxygen (C/O) ratio, largely due to instrumental limitations and the relative faintness of the host star.

The C/O ratio has long been regarded as a key tracer of the formation and migration history of giant exoplanets \citep[e.g.,][]{2011ApJ...743L..16O, 2018A&A...613A..14E, 2019A&A...632A..63C}. Super-solar C/O ratios inferred for several UHJs \citep{2024AJ....168...14W, 2024AJ....168..293S, saha2025a, saha2025b} are consistent with formation beyond the snow line followed by inward migration. However, accretion of inner protoplanetary dust and/or planetesimals is plausible during such migration scenarios and may alter or contaminate the observable atmospheric composition. The C/O ratio has also been proposed as a diagnostic for distinguishing between formation via core accretion and gravitational collapse \citep[e.g.,][]{2014ApJ...794L..12M, 2024ApJ...969L..21B}. In addition, oxygen depletion through sequestration into refractory condensates has been suggested as a mechanism that could artificially elevate the observed C/O ratios in gas-giant atmospheres \citep{2023MNRAS.520.4683F}. Nonetheless, the relatively smaller number of such measurements currently limits the ability to robustly constrain formation pathways.

Robust estimation of the C/O ratio in planetary atmospheres requires precise measurements of the abundances of both carbon- and oxygen-bearing molecules. Since many of these species exhibit strong absorption features in the near- to mid-infrared, earlier ground-based instruments—limited both in sensitivity and spectral coverage in these wavelengths—have struggled to provide the necessary precision. The James Webb Space Telescope \citep[JWST,][]{2006SSRv..123..485G}, with its broad and high-precision spectroscopic coverage across this critical wavelength range, marks a major advancement. It enables the accurate detection of multiple key molecular species, thereby allowing unprecedented precision in the retrieval of atmospheric abundances, C/O ratios, and metallicities.

In this study, we present a comprehensive atmospheric characterization of the day side of the ultra-hot Jupiter WASP-19b, using archival panchromatic JWST emission spectroscopy obtained with NIRSpec/PRISM. Section \ref{sec:meth} describes our methodology, including the data reduction, analyses and atmospheric modeling, while Section \ref{sec:res} presents the key results and discuss their broader implications.

\section{Methodology \label{sec:meth}}

\subsubsection*{Observational data\label{sec:meth1}}

WASP-19b was observed by JWST during cycle 1 as part of the GTO program $\#$1274 (PI: J Lunine). The observation, carried out on February 18, 2023, employed NIRSpec/PRISM \citep{2022A&A...661A..80J} in the Bright Object Time Series (BOTS; \citealp{2023PASP..135a8002E}) mode. This configuration enables low-resolution time-series spectroscopy across a broad wavelength range of $\sim$0.6–5.2$\mu$m. The lower spectral resolution can lead to saturation in moderately bright targets, making it unsuitable for many luminous sources. WASP-19b is relatively faint, which avoids immediate saturation, and allows broad wavelength coverage within a single transit. We accessed the publicly available data from the Barbara A. Mikulski Archive for Space Telescopes (MAST\footnote{https://mast.stsci.edu/}; \dataset[doi: 10.17909/zya5-mg28)]{\doi{10.17909/zya5-mg28}}. The observations span approximately 5.8 hours, covering the entire secondary eclipse of WASP-19b along with a substantial out-of-eclipse baseline. Full coverage of the secondary eclipse is essential for accurately modeling instrumental and astrophysical trends, particularly in high-precision emission spectroscopy.

\subsubsection*{Data reduction\label{sec:meth2}}

The official JWST archive provides data at various calibration levels from the JWST Science Calibration Pipeline \citep{2022zndo...7071140B}, ranging from stage 0 to stage 3. Stage 0 files contain the raw data in FITS format (“*uncal.fits"), which have four dimensions: detector columns, detector rows, the number of groups per integration, and the number of integrations per exposure. The data are divided into multiple sectors (three in our case) to accommodate their large volume. The observation uses 4 groups per integration and a total of 18231 integrations, providing exceptionally high cadence and high signal-to-noise (S/N) for time-series spectroscopy. The PRISM observations use the subarray SUB512, which has a spectral extent of 512 pixels on the NRS1 detector.

We reduced the stage 0 data using two independent pipelines: Eureka! \citep{2022JOSS....7.4503B} and exoTEDRF \citep[formerly supreme-SPOON,][]{2024JOSS....9.6898R}. Both are well-established community pipelines, extensively validated (particularly Eureka!) through several previously published applications on JWST data (e.g., \citealp{2023Natur.614..664A, 2023NatAs...7.1317L, 2024Natur.626..979P, 2023ApJ...959L...9M} for Eureka! and \citealp{2024ApJ...962L..20R, 2023Natur.614..670F} for exoTEDRF). Using two independent reductions provides a cross-validation of the results, an important step in sensitivity-limited studies such as exoplanet atmospheric spectroscopy.

Both pipelines employ modular data reduction workflows that include key calibration and correction procedures—bias subtraction, flat-fielding, ramp fitting, and 1/f noise correction—followed by spectroscopic extraction of the time-series data. Calibration reference files were retrieved from the JWST Calibration Reference Data System (CRDS) \footnote{https://jwst-crds.stsci.edu/}. We closely followed the official documentation of both pipelines, adopting the recommended parameter settings and optimizations from our previous works \citep{saha2025a, saha2025b}. The final extraction resulted in two-dimensional lightcurves resolved in both time and wavelength.

\begin{figure*}
	\centering
	\includegraphics[width=2\columnwidth]{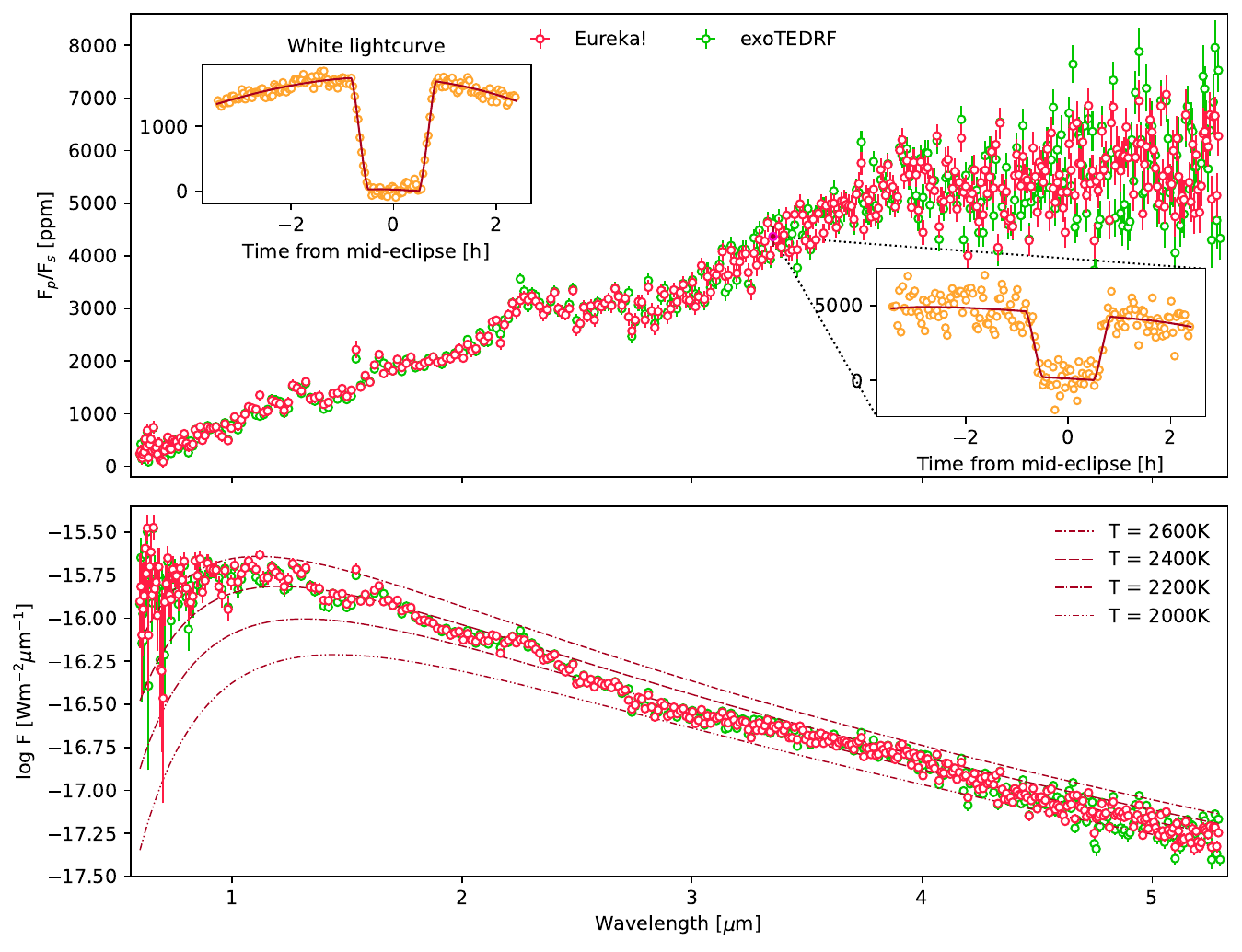}
	\caption{\textit{Top:} Spectroscopic planet-to-star flux ratios with 1$\sigma$ uncertainties, derived from two independent reductions of the NIRSpec/PRISM data using Eureka! and exoTEDRF, followed by the modeling of the spectroscopic secondary eclipse lightcurves. The white lightcurve and the 196th spectroscopic lightcurve (Eureka!, both binned to 2-minute cadence for clarity) are shown along with their respective best-fit secondary-eclipse models. \textit{Bottom:} Planetary emission spectra, normalized to the observer's location, derived from the planet-to-star flux ratios above and modeled PHEONIX stellar spectrum. The mean absolute difference between the two independently reduced spectra is $\sim$1$\sigma$, indicating excellent statistical consistency. For reference, blackbody curves corresponding to 2600K, 2400K, 2200K, and 2000K are also shown. The significant deviation of the planet's spectra from blackbody expectations at shorter wavelengths is due to the dominating contribution of the  reflected star-light by the planet.  \label{fig:fig1}}
\end{figure*}

\subsubsection*{Lightcurves Modeling \label{sec:meth3}}

In contrast to the much higher resolution grisms, the NIRSpec/PRISM observations are much more sparsely resolved in wavelength, and hence common binning practices used for grism observations are not optimal for these data. We have tested various binning schemes to compare the SNR of the lightcurves and the resulting planetary emission spectra, and found that even pixel level lughtcurves have sufficiently high SNR to fit the secondary eclipses precisely. This is partly due to the fact that each PRISM pixel already covers a large enough wavelength range to deliver high-SNR data for relatively bright sources such as the host star WASP-19, and also because UHJs such as WASP-19b have substantially large secondary eclipse depths.

Therefore, we adopted pixel-level spectroscopic lightcurves for our analysis, which resulted in a total of 403 spectroscopic secondary eclipse lightcurves. This yields a significantly high effective resolution that allows us to distinguish between different molecular features, comparable to many grism-based studies in the literature, albeit at somewhat higher uncertainties and increased point-to-point scatter due to the lower overall brightness of the host star. However, because emission spectroscopic features from UHJs are expected to be among the most prominent planetary atmospheric signatures observable to date, the slightly larger uncertainties do not significantly affect our atmospheric interpretation.

To model the 1D lightcurves, we used our in-house pipeline, ExoELF (ExoplanEts Lightcurves Fitter) \citep{2021AJ....162...18S, 2021AJ....162..221S, saha2025a, saha2025b, 2025MNRAS.539..928S}. This pipeline integrates several open-source packages, including batman \citep{2015PASP..127.1161K} for transit and eclipse modeling, emcee \citep{2013PASP..125..306F} and DYNESTY \citep{2020MNRAS.493.3132S} for MCMC and nested sampling, respectively, and celerite \citep{celerite1, celerite2} and George \citep{2015ITPAM..38..252A} for Gaussian-process (GP) regression. ExoELF is optimized to handle large volumes of time-series data from HST and JWST observations, making it suitable for further applications as well.

Comparing the lightcurves from two independent reductions, we find that the exoTEDRF uncertainties remain effectively constant across large pixel ranges but exhibit pronounced, localized fluctuations. This is particularly noticeable in the pixel-level spectroscopic lightcurves (see Figure \ref{fig:figAerr}). Such behavior likely originates from computational approximations at certain stages of the reduction process, which also caused exoTEDRF uncertainties to be somewhat unreliable relative to those from Eureka!. In the white lightcurves, the fluctuations strongly dominate the exoTEDRF uncertainties (see Figure \ref{fig:figAerr}). In contrast, the Eureka uncertainties appear consistent with the observed scatter in both spectroscopic and white lightcurves. For these reasons, we consider the Eureka! reduction to be more reliable and adopt it as the default dataset for our primary results.

We implemented a two-stage procedure to fit the lightcurves. In the first stage, we fit the white lightcurve with the mid-transit time, orbital parameters (scaled orbital semi-major axis and orbital inclination), eclipse depth, and detrending parameters free (see Table \ref{tab:tr_fit}). In the second stage, we fit the spectroscopic lightcurves while fixing the mid-transit time and orbital parameters to the values obtained from the white lightcurve fit. This approach avoids poor constraints on wavelength-independent parameters in the lower-SNR spectroscopic lightcurves.

Proper detrending during lightcurve fitting is crucial to accurately estimate the secondary eclipse depths. We tested various detrending functions, from simple polynomials to GP regression. Because our lightcurves are mostly flat with only long-term trends, and the secondary eclipse signals have smaller amplitudes relative to the lightcurve SNRs—especially in spectroscopic lightcurves—complex detrending methods like GP regression tend to overfit and can misinterpret eclipse signals as trends. Similarly, very high-order polynomials can overfit the data. After testing several lower-order polynomials, we found that a fourth-order polynomial provides the best compromise, capturing both symmetric and asymmetric long-term trends. To ensure numerical stability, we used a fourth-order Legendre polynomial expansion instead of a standard polynomial, whose orthogonality mitigates correlations between higher-order terms.

Our combined lightcurve fits were performed using a nested sampling algorithm in DYNESTY. The emission spectra obtained from both independent reductions (see Figure \ref{fig:fig1}) have a mean absolute difference of $\sim$1$\sigma$, indicating excellent statistical agreement.

\begin{figure*}
	\centering
	\includegraphics[width=2\columnwidth]{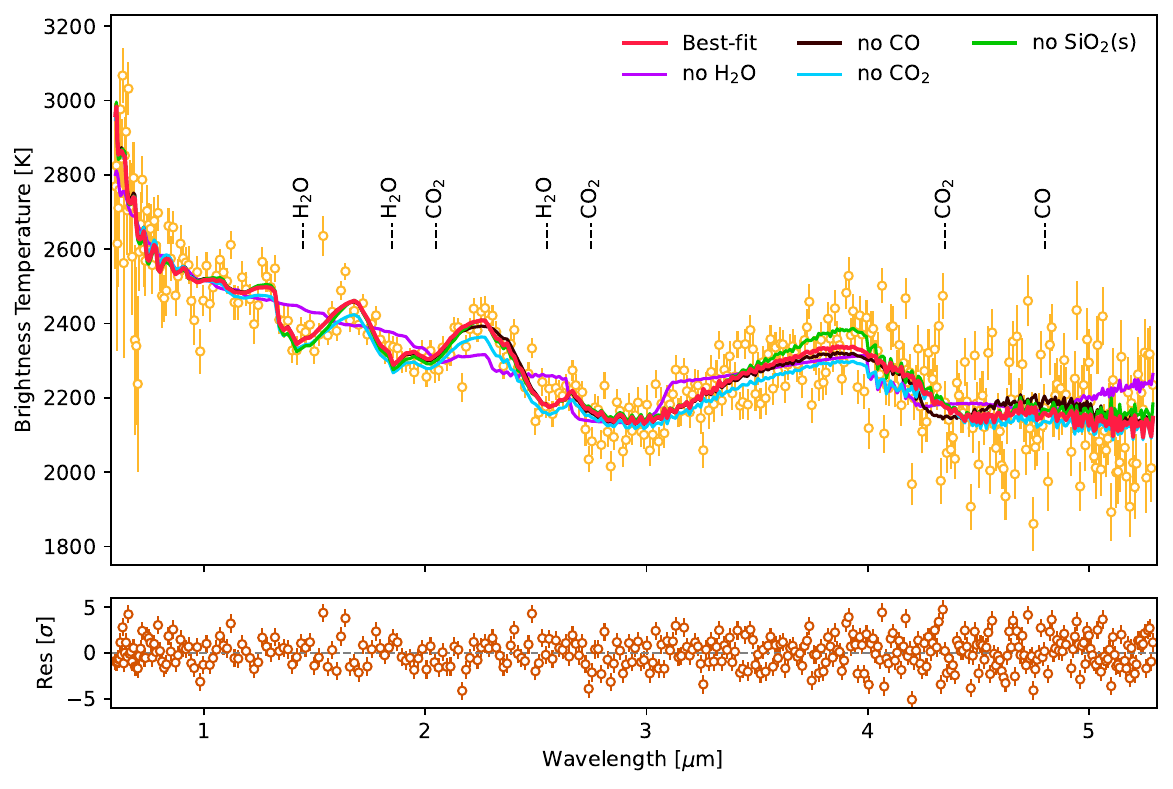}
	\caption{The observed emission spectrum of WASP-19b in terms of brightness temperatures (orange, Eureka!) is shown, along with the best-fit free-chemistry retrieved model and the residuals. Retrieved models excluding individual species are also shown, which were used to assess the statistical significance of their detection. Key spectral features from H$_2$O, CO, and CO$_2$, each detected with strong statistical significance, are also highlighted for reference. \label{fig:fig2}}
\end{figure*}

\subsubsection*{Atmospheric retrievals \label{sec:meth4}}

Thanks to the high S/N of our extracted emission spectra, we performed a detailed atmospheric characterization using retrieval analyses. For this, we adopted petitRADTRANS \citep[pRT,][]{2019A&A...627A..67M}, a widely used and well-validated radiative transfer code in the exoplanet community \citep[e.g.,][]{2024arXiv241008116K, 2024ApJ...973L..41I, 2024arXiv241203675L, 2024MNRAS.527.7079P, 2024A&A...690A..63B}. While pRT employs a one-dimensional radiative transfer framework, it is robust enough for the reliable interpretation of exoplanet spectra from current instruments such as JWST, offering nearly all functionalities needed for complex atmospheric analyses.

Our retrieval framework is based on the pRT retrieval routines \citep{2024JOSS....9.5875N}, which use PyMultinest \citep{2009MNRAS.398.1601F, 2014A&A...564A.125B} for nested sampling of the posterior space. Given the high spectral resolution of our data, we adopted the line-by-line opacity mode with a model resolution of 20000, appropriate for such retrievals. The pressure-temperature (P–T) profile was parametrized following \citet{2010A&A...520A..27G}, which provides sufficient flexibility without excessive computational cost. This parametrization involves four variables: the equilibrium (irradiation) temperature (T$_{\mathrm{eq}}$), the internal temperature (T$_{\mathrm{int}}$), the visible-to-infrared opacity ratio ($\gamma$), and the mean infrared opacity ($\kappa_{\mathrm{IR}}$).

We applied both free and equilibrium chemistry models to describe the chemical abundances. In the free chemistry approach, the abundances of individual species (expressed as log-MMRs) were treated as vertically uniform free parameters, typically ranging between -14 and -0.2 (see Table \ref{tab:retrieval_priors}). To improve computational efficiency, the upper bound was lowered for species not requiring a higher range in preliminary fits. For equilibrium chemistry, we used precomputed abundance tables from \texttt{pRT}, assuming chemical equilibrium at each pressure level.

Our models included opacities of all major molecules expected to contribute significantly across the observed wavelengths, including H$_2$O \citep{2018MNRAS.480.2597P}, CO \citep{2010JQSRT.111.2139R}, CO$_2$ \citep{2010JQSRT.111.2139R}, SiO \citep{2013MNRAS.434.1469B}, CH$_4$ \citep{2020ApJS..247...55H}, HCN \citep{2006MNRAS.367..400H}, C$_2$H$_2$ \citep{2013JQSRT.130....4R}, PH$_3$ \citep{2015MNRAS.446.2337S}, H$_2$S \citep{2013JQSRT.130....4R}, TiO \citep{2019A&A...627A..67M}, and VO \citep{2019A&A...627A..67M}. Rayleigh scattering by H$_2$ \citep{1962ApJ...136..690D} and He \citep{1965PPS....85..227C} and collision-induced absorption (CIA) from H$_2$-H$_2$ \citep{2001JQSRT..68..235B, 2002A&A...390..779B} and H$_2$-He \citep{1988ApJ...326..509B, 1989ApJ...336..495B} were included. Bound-free and free-free absorption by H$^-$ \citep{2008oasp.book.....G} was also considered, as it can be significant in ultra-hot Jupiter atmospheres.

We incorporated several condensate species expected to form clouds at these temperatures, following the formalism by \citet{2001ApJ...556..872A}, including MgSiO$_3$(s) \citep{1994A&A...292..641J, 2001ApJ...556..872A}, Mg$_2$SiO$_4$(s) \citep{1973PSSBR..55..677S, 2017Icar..289...42C}, Al$_2$O$_3$(s) \citep{1995Icar..114..203K, Stull1947}, SiO$_2$(s) \citep{2018MNRAS.475...94K, 1997A&A...327..743H, 1995Icar..114..203K, Stull1947}, TiO$_2$(s) \citep{2018MNRAS.475...94K, 2011A&A...526A..68Z, 2003ApJS..149..437P, 2016arXiv160704866S, 2019RJPCA..93.1024S}, and CaTiO$_3$(s) \citep{2018MNRAS.475...94K, 2003ApJS..149..437P, 1998JPCM...10.3669U, 2019RJPCA..93.1024S}. Their abundances at the cloud base were treated as free parameters in both free and equilibrium chemistry models. Additionally, we treated the sedimentation efficiency (f${\mathrm{sed}}$), the log-normal width of the particle size distribution ($\sigma$), the vertical eddy diffusion coefficient (k${\mathrm{zz}}$), and the cloud fraction (f${\mathrm{c}}$) as free parameters, which together constrain the cloud profiles \citep{2001ApJ...556..872A}. Finally, we included dayside scattering of starlight using a dayside averaged irradiance.

\begin{table*}
	\centering
	\caption{Molecular abundances (in log-MMRs) and log-evidence from the atmospheric retrievals.}
	\label{tab:tab1}
	\begin{tabular}{lccccccc}
		\hline
		Model & [H$_2$O] & [CO] & [CO$_2$] & [SiO] & [TiO] & [VO] & $\ln Z$ \\
		\hline
		Eureka! (free chem.)\\
		Best-fit & $-3.51_{-0.13}^{+0.18}$ & $-2.8_{-0.33}^{+0.38}$ & $-6.2_{-0.24}^{+0.2}$ & $-4.21_{-1.72}^{+0.72}$ & $-11.68_{-1.3}^{+1.25}$ & $-8.92_{-0.24}^{+0.21}$ & 15535.29 \\
		No H$_2$O & – & $-0.83_{-0.08}^{+0.07}$ & $-2.9_{-0.08}^{+0.09}$ & $-0.77_{-0.1}^{+0.07}$ & $-9.0_{-0.54}^{+0.7}$ & $-8.7_{-2.22}^{+0.5}$ & 15327.07 \\
		No CO & $-3.69_{-0.09}^{+0.12}$ & – & $-5.95_{-0.11}^{+0.14}$ & $-8.38_{-2.09}^{+2.18}$ & $-11.41_{-1.37}^{+1.23}$ & $-9.02_{-0.39}^{+0.27}$ & 15523.80 \\
        No CO$_2$ & $-3.41_{-0.14}^{+0.15}$ & $-2.37_{-0.27}^{+0.29}$ & – & $-3.04_{-0.31}^{+0.33}$ & $-11.86_{-1.29}^{+1.22}$ & $-8.85_{-0.21}^{+0.17}$ & 15533.06 \\
        No SiO & $-3.5_{-0.14}^{+0.17}$ & $-2.83_{-0.35}^{+0.34}$ & $-6.1_{-0.18}^{+0.21}$ & – & $-11.73_{-1.22}^{+1.24}$ & $-8.89_{-0.29}^{+0.21}$ & 15537.23 \\
        No TiO & $-3.52_{-0.13}^{+0.15}$ & $-2.83_{-0.3}^{+0.32}$ & $-6.21_{-0.25}^{+0.19}$ & $-4.26_{-1.05}^{+0.69}$ & – & $-8.89_{-0.17}^{+0.17}$ & 15535.93 \\
        No VO & $-3.32_{-0.15}^{+0.17}$ & $-2.37_{-0.28}^{+0.28}$ & $-6.03_{-0.28}^{+0.22}$ & $-3.65_{-0.51}^{+0.47}$ & $-9.6_{-0.18}^{+0.19}$ & – & 15531.33 \\
        No clouds & $-3.45_{-0.13}^{+0.18}$ & $-2.51_{-0.28}^{+0.35}$ & $-6.33_{-1.3}^{+0.3}$ & $-2.96_{-0.46}^{+0.45}$ & $-10.84_{-2.12}^{+0.96}$ & $-9.17_{-1.14}^{+0.24}$ & 15526.06 \\
        No SiO$_2$(s) & $-3.46_{-0.13}^{+0.17}$ & $-2.49_{-0.28}^{+0.34}$ & $-6.35_{-1.48}^{+0.3}$ & $-2.95_{-0.42}^{+0.47}$ & $-11.23_{-1.85}^{+1.22}$ & $-9.15_{-0.73}^{+0.22}$ & 15527.58 \\
		\hline
		Eureka! (eq. chem.)\\
		Best-fit & $-3.9_{-0.04}^{+0.04}$ & $-2.29_{-0.02}^{+0.03}$ & $-7.84_{-0.07}^{+0.07}$ & $-3.74_{-0.04}^{+0.05}$ & $-6.86_{-0.05}^{+0.05}$ & $-9.21_{-0.05}^{+0.06}$ & 15528.51 \\
        No clouds & $-3.85_{-0.04}^{+0.05}$ & $-2.37_{-0.03}^{+0.04}$ & $-7.99_{-0.08}^{+0.09}$ & $-3.75_{-0.03}^{+0.04}$ & $-6.61_{-0.04}^{+0.05}$ & $-8.88_{-0.05}^{+0.06}$ & 15497.91 \\
		\hline
		exoTEDRF (free chem.)\\
		Best-fit & $-3.32_{-0.14}^{+0.15}$ & $-2.59_{-0.27}^{+0.28}$ & $-5.99_{-0.24}^{+0.2}$ & $-3.78_{-0.71}^{+0.62}$ & $-9.46_{-0.37}^{+0.51}$ & $-9.78_{-1.8}^{+0.79}$ & 15500.30 \\
        No H$_2$O & – & $-0.72_{-0.15}^{+0.13}$ & $-2.37_{-0.12}^{+0.1}$ & $-1.16_{-0.43}^{+0.28}$ & $-8.34_{-1.5}^{+0.8}$ & $-8.25_{-2.65}^{+0.74}$ & 15500.30 \\
        No CO & $-3.14_{-0.27}^{+0.24}$ & – & $-5.47_{-0.25}^{+0.23}$ & $-7.84_{-2.25}^{+2.23}$ & $-9.43_{-1.14}^{+0.8}$ & $-9.13_{-0.73}^{+0.56}$ & 15493.46 \\
        No CO$_2$ & $-3.23_{-0.15}^{+0.19}$ & $-2.12_{-0.29}^{+0.35}$ & – & $-2.33_{-0.44}^{+0.5}$ & $-9.69_{-0.31}^{+0.41}$ & $-10.02_{-2.03}^{+0.85}$ & 15496.93 \\
        No SiO & $-3.24_{-0.16}^{+0.21}$ & $-2.56_{-0.32}^{+0.34}$ & $-5.81_{-0.18}^{+0.22}$ & – & $-9.45_{-0.68}^{+0.68}$ & $-9.24_{-0.92}^{+0.57}$ & 15502.49 \\
        No TiO & $-3.21_{-0.16}^{+0.18}$ & $-2.39_{-0.3}^{+0.34}$ & $-5.9_{-0.32}^{+0.22}$ & $-3.54_{-0.68}^{+0.66}$ & – & $-8.65_{-0.26}^{+0.3}$ & 15500.87 \\
		No VO & $-3.28_{-0.14}^{+0.15}$ & $-2.45_{-0.29}^{+0.33}$ & $-6.01_{-0.35}^{+0.21}$ & $-3.41_{-0.59}^{+0.69}$ & $-9.54_{-0.21}^{+0.29}$ & – & 15498.65 \\
        No clouds & $-3.23_{-0.17}^{+0.23}$ & $-2.05_{-0.35}^{+0.44}$ & $-6.45_{-3.25}^{+0.51}$ & $-2.17_{-0.52}^{+0.55}$ & $-9.92_{-0.7}^{+0.21}$ & $-9.99_{-2.39}^{+0.85}$ & 15494.38 \\
        No SiO$_2$(s) & $-3.25_{-0.15}^{+0.2}$ & $-2.08_{-0.32}^{+0.4}$ & $-6.41_{-3.18}^{+0.44}$ & $-2.22_{-0.49}^{+0.52}$ & $-9.9_{-0.46}^{+0.2}$ & $-10.25_{-2.26}^{+0.88}$ & 15496.16 \\
		\hline
		exoTEDRF (eq. chem.)\\
		Best-fit & $-3.83_{-0.05}^{+0.04}$ & $-2.26_{-0.03}^{+0.05}$ & $-7.74_{-0.07}^{+0.08}$ & $-3.64_{-0.08}^{+0.1}$ & $-6.79_{-0.06}^{+0.07}$ & $-9.14_{-0.06}^{+0.07}$ & 15493.25 \\
        No clouds & $-3.83_{-0.04}^{+0.05}$ & $-2.35_{-0.03}^{+0.04}$ & $-7.95_{-0.08}^{+0.09}$ & $-3.73_{-0.03}^{+0.04}$ & $-6.58_{-0.04}^{+0.05}$ & $-8.84_{-0.05}^{+0.07}$ & 15455.61 \\
		\hline
	\end{tabular}
\end{table*}

\section{Results and Discussion \label{sec:res}}

The lower resolution of NIRSpec/PRISM enables high S/N observations for comparatively faint targets while providing exceptionally broad wavelength coverage $\sim$0.6–5.2$\mu$m from a single time-series observation. Moreover, due to their elevated temperatures and inflated radii, UHJs exhibit much stronger emission features than cooler exoplanets. Together, these factors make the emission spectra of WASP-19b ideally suited for precision retrieval analyses.

The free chemistry retrieval of the Eureka! spectrum yielded tightly constrained abundances of H$_2$O, CO, and CO$_2$, along with marginally constrained abundances for SiO, TiO, and VO (see Table \ref{tab:tab1}, Figure \ref{fig:fig2} and \ref{fig:figA2}). The retrieval also placed strong constraints on the cloud properties and the base abundance of SiO$_2$(s), indicating its presence. To assess the statistical significance \citep{Kass01061995} of each molecular detection, we performed additional retrievals excluding individual species from the model one at a time (see Table \ref{tab:tab1}, Figure \ref{fig:fig2}). From these comparisons, we find significant statistical evidence for the detection of H$_2$O (20.41$\sigma$) and CO (4.79$\sigma$), as well as marginal evidence for CO$_2$ (2.11$\sigma$) and VO (2.81$\sigma$). We found no statistically significant detection of SiO, although its abundance is well constrained from the retrievals. We also find no statistically significant detection of TiO despite a tight upper limit. We also find no evidence for contributions from H$^-$ bound-free and free-free absorption opacities.

\begin{figure*}
	\centering
    \begin{tabular}{cc}
	\includegraphics[width=\columnwidth]{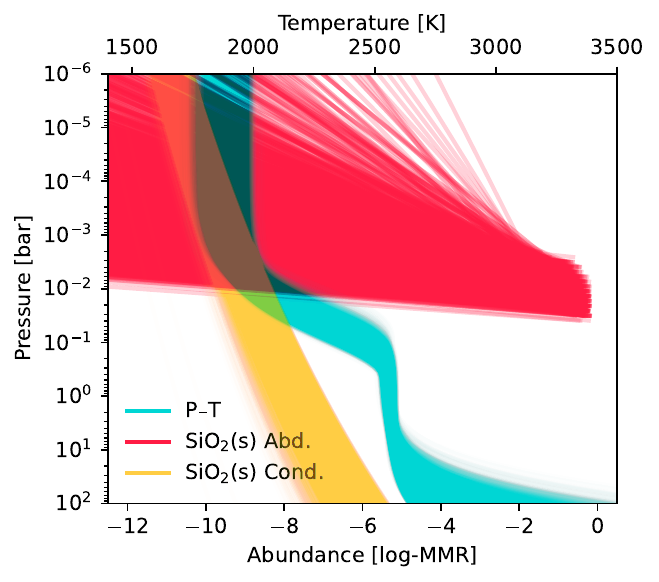} & \includegraphics[width=\columnwidth]{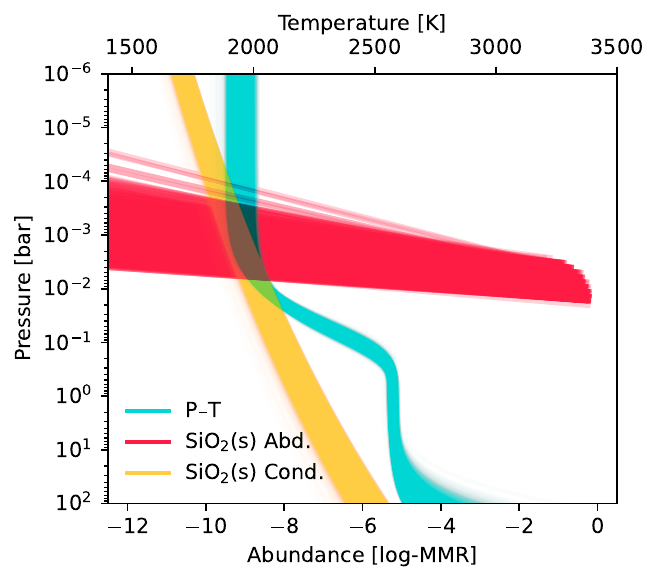}
    \end{tabular}
	\caption{The retrieved posterior P–T profiles from the free (left) and equilibrium chemistry (right) retrievals of the Eureka! spectrum are shown, along with the corresponding SiO$_2$(s) condensation curves and their abundance profiles. \label{fig:fig3}}
\end{figure*}

\begin{figure}
	\centering
	\includegraphics[width=\columnwidth]{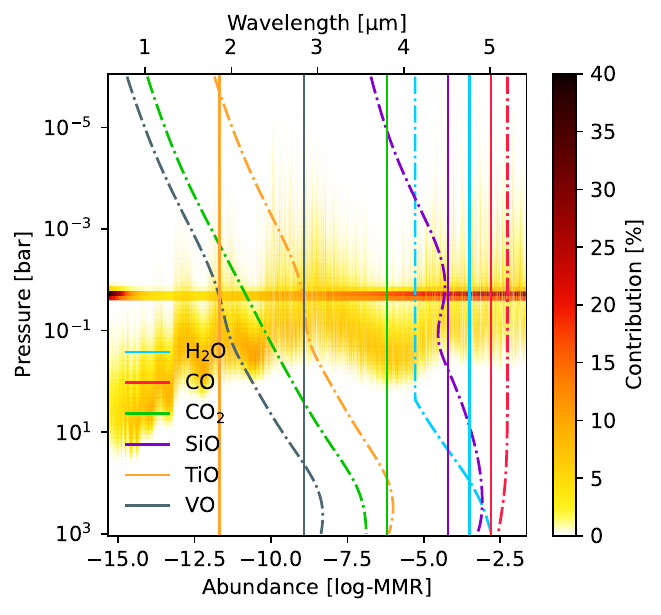}
	\caption{The retrieved median abundances of H$_2$O, CO, CO$_2$, SiO, TiO, and VO are shown from both free (solid) and equilibrium (dashed-dotted) retrievals for comparison (Eureka!), along with the median contribution function from the free chemistry retrievals. The strong contribution from SiO$_2$(s) cloud is apparent. \label{fig:fig4}}
\end{figure}

We also report statistically significant evidence for the presence of clouds (4.3$\sigma$), with strong evidence for silicon dioxide (also called silica/quartz, SiO$_2$(s)) condensates (3.93$\sigma$, see Table \ref{tab:tab1} and Figure \ref{fig:fig3}). The contribution function shows strong features arising from the reflected star-light from the clouds at the shorter visible wavelengths, as well as gradually increasing emission features towards the longer wavelength end of the spectrum (see Figure \ref{fig:fig4}). This makes WASP-19b the first UHJ with a statistically significant detection of clouds. We note that a parallel detection of condensate clouds (CaTiO$_3$(s)) has been reported by our team for WASP-121b \citep{saha2025a}, which can be considered as the second such detection, as also discussed in that work.

To independently verify these findings, we also performed analogous retrievals on the exoTEDRF spectrum, which yielded statistically consistent constraints on the molecular abundances (see Table \ref{tab:tab1} and Figure \ref{fig:figA5}). Retrievals excluding specific species indicated significant statistical evidence for H$_2$O (21.29$\sigma$) and CO (3.7$\sigma$), along with marginal evidence for CO$_2$ (2.6$\sigma$) and VO (1.82$\sigma$), very similar to those obtained from the analysis of the Eureka! spectrum. Similarly, we find no statistically significant detection for SiO or TiO. The analysis also showed a consistent detection of clouds (3.44$\sigma$), with strong evidence for SiO$_2$(s) condensates (2.88$\sigma$).

Performing the equilibrium chemistry retrievals on the Eureka! spectrum, we found that the constrained molecular abundances were broadly consistent with those from the free chemistry retrievals (see Table \ref{tab:tab1}, Figure \ref{fig:fig4}, \ref{fig:figA2eq}, and Figure \ref{fig:figA4}). The only notable changes were a slightly elevated CO abundance and a much higher TiO abundance. The elevated CO abundance likely arises from the model attempting to maintain CO/H$_2$O ratios consistent with the pre-calculated abundance tables under chemical equilibrium. Similarly, the significantly higher TiO abundance likely arises from the pre-calculated equilibrium abundances expected at the retrieved temperatures, together with the well-constrained abundances of dominant species such as H$_2$O. While this may suggest a potential depletion of TiO relative to the expected equilibrium abundances at the retrieved temperatures, the absence of robust TiO detection renders this inference inconclusive.

It is also worth noting that the overall abundances of the molecular species and the pressure-temperature (P–T) profiles retrieved using the equilibrium chemistry models appear more precisely constrained, which is mainly due to the additional constraints imposed by the priors derived from the pre-computed tables (see Table \ref{tab:tab1} and Figure \ref{fig:fig3}), and may therefore not fully reflect the true atmospheric conditions. This is further reflected in the fact that the retrieved best-fit model from the equilibrium chemistry retrieval is statistically 3.68 $\sigma$ less significant than the free chemistry fit. Nonetheless, the equilibrium chemistry fit also indicated strong evidence for the presence of clouds (7.82 $\sigma$), further supporting our inference of clouds in this dayside atmosphere.

\begin{figure}
	\centering
	\includegraphics[width=1\columnwidth]{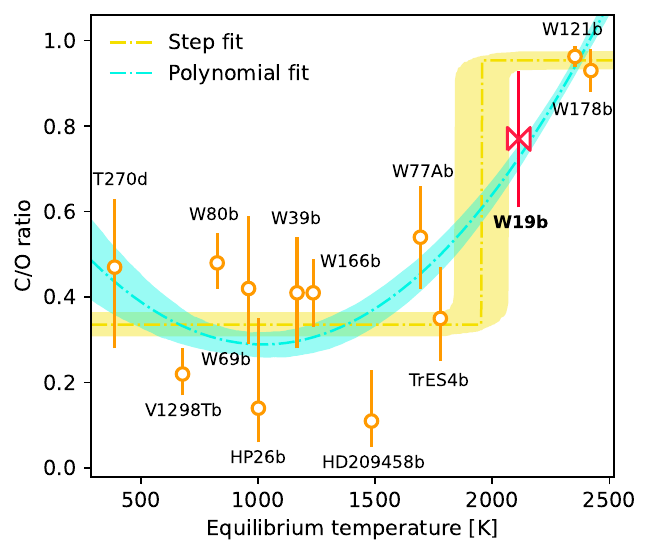}
	\caption{Several reported C/O ratios for a wide range of exoplanets based on JWST observations are shown, plotted against their equilibrium temperatures, with the estimated C/O ratio for WASP-19b from this work highlighted as the magenta bowtie. A step-function model illustrating the potential onset of super-solar C/O ratios in UHJs is shown, along with a quadratic polynomial model representing a gradual increase of the C/O ratio with equilibrium temperature, both with their corresponding 1$\sigma$ uncertainties. \label{fig:figco}}
\end{figure}

Leveraging the precisely constrained abundances of the dominant molecular species, which show statistically significant detections (H$_2$O, CO, and CO$_2$) from the free chemistry retrievals of the Eureka! spectrum, we derive a highly constrained atmospheric C/O ratio of 0.77 $\pm$ 0.16. When we also include the retrieved abundance of SiO, which remains a statistical non-detection, we obtain a C/O ratio of 0.75 $\pm$ 0.17, which is highly consistent with the former value. Similarly, using the retrieved abundances from the free chemistry modeling of the exoTEDRF spectrum, we obtained a C/O ratio of 0.78 $\pm$ 0.12 (0.75 $\pm$ 0.14 when including SiO), which is likewise highly consistent. This potentially super-solar C/O ratio in the dayside atmosphere of WASP-19b is consistent with the emerging trend of elevated C/O ratios reported in the recent studies of other UHJs using both JWST and non-JWST observations.

Given JWST's capability to tightly constrain key molecular abundances, we compiled a literature sample of precisely measured C/O ratios for a broad population of exoplanets to examine possible population-level trends (Figure~\ref{fig:figco}). The literature values were adopted from \citep{2024arXiv240303325B, 2024AJ....168...14W, 2024ApJ...963L...5X, 2024ApJ...962L..30E, 2011Natur.469...64M, 2024A&A...685A..64K, 2025arXiv250601800W, 2025MNRAS.539.1381M, 2025AJ....170..292G, 2025arXiv251027223D, 2025AJ....170..165B, 2025AJ....170...50M, saha2025a, saha2025b}, and several of the corresponding equilibrium temperatures were taken from \citep{2023ApJS..268....2S, 2024ApJS..274...13S}. 

To investigate the potential onset of super-solar C/O ratios among the hottest exoplanets, we modeled the compiled C/O ratios using a step function fit (see Figure \ref{fig:figco}), obtaining C/O$_{\mathrm{low}}$ = 0.34 $\pm$ 0.03, C/O$_{\mathrm{high}}$ = 0.95 $\pm$ 0.02, and T$_{\mathrm{break}}$ = 1953.2 $\pm$ 111.5 K. We also tested a scenario in which the C/O ratio gradually increases with equilibrium temperature by modeling the compiled C/O ratios using a quadratic polynomial function, C/O = a$_0$ + a$_1$T${_\mathrm{eq}}$ + a$_2$T${_\mathrm{eq}}^2$ (see Figure \ref{fig:figco}), yielding a$_0$ = 0.67 $\pm$ 0.16, a$_1$ = -7.47 × 10$^{-4}$ $\pm$ 2.38 × 10$^{-4}$, and a$_2$ = 3.66 × 10$^{-7}$ $\pm$ 7.43 × 10$^{-8}$.

Comparing both fits, we find that the step function fit provides a slightly better fit for the data than the quadratic polynomial model with $\Delta$BIC = 2.75. Although this evidence is weak, the comparison highlights a potential onset of highly elevated C/O ratios in UHJs. While the current dataset remains limited and many reported values have yet to be independently verified, these results suggest a possible transition toward super-solar C/O ratios at $\mathrm{T_{eq}} \gtrsim 2000$ K. We emphasize, however, that a larger sample of uniformly analyzed targets will be essential to confirm and refine this apparent trend. In particular, obtaining precise C/O measurements for additional UHJs and hot Jupiters near the transition region ($\sim$2000 K) will be critical for establishing the robustness of this trend and for improving our understanding of the underlying physical mechanisms.

Although super-solar C/O ratios in exoplanet atmospheres have previously been attributed to formation beyond the ice line followed by inward migration \citep[e.g.,][]{2017AJ....153...83B, 2021A&A...654A..71S}, this mechanism alone cannot account for the highly elevated C/O values observed in UHJs. Many hot Jupiters are also believed to have formed at wide separations before migrating inward, yet they exhibit substantially lower C/O ratios. Instead, we hypothesize that the onset of extreme C/O ratios in UHJs originates from the sequestration of oxygen-bearing condensates on their nightsides \citep[e.g.,][]{2023MNRAS.520.4683F}. The extremely high temperatures in UHJ dayside atmospheres act as a catalyst for this process by maintaining refractory species such as TiO and SiO in gaseous form. As these gases cool in the upper atmosphere, they condense into refractory cloud particles—such as SiO$_2$(s) in WASP-19b and CaTiO$_3$(s) in WASP-121b—which may then be advected to the nightside by strong atmospheric circulation. There, these condensates can rain or become cold-trapped, ultimately depleting the dayside of oxygen and leading to the observed enhancement in the C/O ratio.

Using the estimated molecular abundances from the free chemistry retrieval of the Eureka! spectrum, we inferred a metallicity of 0.2$_{-0.1}^{+0.28}$ $\times$solar. Although this value is sub-solar, the estimate is strongly driven by the most dominant molecule, CO, and therefore primarily reflects a slightly depleted CO abundance. This is also apparent when comparing the retrieved abundances with those from the equilibrium chemistry fit, and could be due to an overall depletion of oxygen in the dayside atmosphere. While any inference regarding formation mechanisms remains inconclusive based on this highly approximated metallicity, scenarios involving substantial mass loss due to photoevaporation can likely be ruled out.

We thank the anonymous reviewer for their constructive comments and suggestions, which have helped improve the manuscript. SS acknowledges Fondo Comité Mixto-ESO Chile ORP 025/2022 to support this research. The computations presented in this work were performed using the Geryon-3 supercomputing cluster, which was assembled and maintained using funds provided by the ANID-BASAL Center FB210003, Center for Astrophysics and Associated Technologies, CATA. This work is based [in part] on observations made with the NASA/ESA/CSA James Webb Space Telescope. The data were obtained from the Mikulski Archive for Space Telescopes at the Space Telescope Science Institute, which is operated by the Association of Universities for Research in Astronomy, Inc., under NASA contract NAS 5-03127 for JWST. These observations are associated with program $\#$2055.  JSJ gratefully acknowledges support by FONDECYT grant 1240738 and from the ANID BASAL project FB210003.

\bibliography{ms}

\begin{thebibliography}{}
\expandafter\ifx\csname natexlab\endcsname\relax\def\natexlab#1{#1}\fi
\providecommand{\url}[1]{\href{#1}{#1}}
\providecommand{\dodoi}[1]{doi:~\href{http://doi.org/#1}{\nolinkurl{#1}}}
\providecommand{\doeprint}[1]{\href{http://ascl.net/#1}{\nolinkurl{http://ascl.net/#1}}}
\providecommand{\doarXiv}[1]{\href{https://arxiv.org/abs/#1}{\nolinkurl{https://arxiv.org/abs/#1}}}

\bibitem[{{Ackerman} \& {Marley}(2001)}]{2001ApJ...556..872A}
{Ackerman}, A.~S., \& {Marley}, M.~S. 2001, \apj, 556, 872, \dodoi{10.1086/321540}

\bibitem[{{Alderson} {et~al.}(2023){Alderson}, {Wakeford}, {Alam}, {Batalha}, {Lothringer}, {Adams Redai}, {Barat}, {Brande}, {Damiano}, {Daylan}, {Espinoza}, {Flagg}, {Goyal}, {Grant}, {Hu}, {Inglis}, {Lee}, {Mikal-Evans}, {Ramos-Rosado}, {Roy}, {Wallack}, {Batalha}, {Bean}, {Benneke}, {Berta-Thompson}, {Carter}, {Changeat}, {Col{\'o}n}, {Crossfield}, {D{\'e}sert}, {Foreman-Mackey}, {Gibson}, {Kreidberg}, {Line}, {L{\'o}pez-Morales}, {Molaverdikhani}, {Moran}, {Morello}, {Moses}, {Mukherjee}, {Schlawin}, {Sing}, {Stevenson}, {Taylor}, {Aggarwal}, {Ahrer}, {Allen}, {Barstow}, {Bell}, {Blecic}, {Casewell}, {Chubb}, {Crouzet}, {Cubillos}, {Decin}, {Feinstein}, {Fortney}, {Harrington}, {Heng}, {Iro}, {Kempton}, {Kirk}, {Knutson}, {Krick}, {Leconte}, {Lendl}, {MacDonald}, {Mancini}, {Mansfield}, {May}, {Mayne}, {Miguel}, {Nikolov}, {Ohno}, {Palle}, {Parmentier}, {Petit dit de la Roche}, {Piaulet}, {Powell}, {Rackham}, {Redfield}, {Rogers}, {Rustamkulov}, {Tan}, {Tremblin}, {Tsai}, {Turner}, {de Val-Borro},
  {Venot}, {Welbanks}, {Wheatley}, \& {Zhang}}]{2023Natur.614..664A}
{Alderson}, L., {Wakeford}, H.~R., {Alam}, M.~K., {et~al.} 2023, \nat, 614, 664, \dodoi{10.1038/s41586-022-05591-3}

\bibitem[{{Ambikasaran} {et~al.}(2015){Ambikasaran}, {Foreman-Mackey}, {Greengard}, {Hogg}, \& {O'Neil}}]{2015ITPAM..38..252A}
{Ambikasaran}, S., {Foreman-Mackey}, D., {Greengard}, L., {Hogg}, D.~W., \& {O'Neil}, M. 2015, IEEE Transactions on Pattern Analysis and Machine Intelligence, 38, 252, \dodoi{10.1109/TPAMI.2015.2448083}

\bibitem[{{Anderson} {et~al.}(2013){Anderson}, {Smith}, {Madhusudhan}, {Wheatley}, {Collier Cameron}, {Hellier}, {Campo}, {Gillon}, {Harrington}, {Maxted}, {Pollacco}, {Queloz}, {Smalley}, {Triaud}, \& {West}}]{2013MNRAS.430.3422A}
{Anderson}, D.~R., {Smith}, A.~M.~S., {Madhusudhan}, N., {et~al.} 2013, \mnras, 430, 3422, \dodoi{10.1093/mnras/stt140}

\bibitem[{{Barat} {et~al.}(2025){Barat}, {D{\'e}sert}, {Mukherjee}, {Goyal}, {Xue}, {Kawashima}, {Vazan}, {Misener}, {Schlichting}, {Fortney}, {Bean}, {Avarsekar}, {Henry}, {Baeyens}, {Line}, {Livingston}, {David}, {Petigura}, {Sikora}, {Shivkumar}, {Feinstein}, \& {Oklop{\v{c}}i{\'c}}}]{2025AJ....170..165B}
{Barat}, S., {D{\'e}sert}, J.-M., {Mukherjee}, S., {et~al.} 2025, \aj, 170, 165, \dodoi{10.3847/1538-3881/adec89}

\bibitem[{{Barton} {et~al.}(2013){Barton}, {Yurchenko}, \& {Tennyson}}]{2013MNRAS.434.1469B}
{Barton}, E.~J., {Yurchenko}, S.~N., \& {Tennyson}, J. 2013, \mnras, 434, 1469, \dodoi{10.1093/mnras/stt1105}

\bibitem[{{Bell} {et~al.}(2022){Bell}, {Ahrer}, {Brande}, {Carter}, {Feinstein}, {Caloca}, {Mansfield}, {Zieba}, {Piaulet}, {Benneke}, {Filippazzo}, {May}, {Roy}, {Kreidberg}, \& {Stevenson}}]{2022JOSS....7.4503B}
{Bell}, T., {Ahrer}, E.-M., {Brande}, J., {et~al.} 2022, The Journal of Open Source Software, 7, 4503, \dodoi{10.21105/joss.04503}

\bibitem[{{Benneke} {et~al.}(2024){Benneke}, {Roy}, {Coulombe}, {Radica}, {Piaulet}, {Ahrer}, {Pierrehumbert}, {Krissansen-Totton}, {Schlichting}, {Hu}, {Yang}, {Christie}, {Thorngren}, {Young}, {Pelletier}, {Knutson}, {Miguel}, {Evans-Soma}, {Dorn}, {Gagnebin}, {Fortney}, {Komacek}, {MacDonald}, {Raul}, {Cloutier}, {Acuna}, {Lafreni{\`e}re}, {Cadieux}, {Doyon}, {Welbanks}, \& {Allart}}]{2024arXiv240303325B}
{Benneke}, B., {Roy}, P.-A., {Coulombe}, L.-P., {et~al.} 2024, arXiv e-prints, arXiv:2403.03325, \dodoi{10.48550/arXiv.2403.03325}

\bibitem[{{Bergin} {et~al.}(2024){Bergin}, {Booth}, {Colmenares}, \& {Ilee}}]{2024ApJ...969L..21B}
{Bergin}, E.~A., {Booth}, R.~A., {Colmenares}, M.~J., \& {Ilee}, J.~D. 2024, \apjl, 969, L21, \dodoi{10.3847/2041-8213/ad5839}

\bibitem[{{Blain} {et~al.}(2024){Blain}, {Landman}, {Molli{\`e}re}, \& {Dittmann}}]{2024A&A...690A..63B}
{Blain}, D., {Landman}, R., {Molli{\`e}re}, P., \& {Dittmann}, J. 2024, \aap, 690, A63, \dodoi{10.1051/0004-6361/202450767}

\bibitem[{{Borysow}(2002)}]{2002A&A...390..779B}
{Borysow}, A. 2002, \aap, 390, 779, \dodoi{10.1051/0004-6361:20020555}

\bibitem[{{Borysow} {et~al.}(1989){Borysow}, {Frommhold}, \& {Moraldi}}]{1989ApJ...336..495B}
{Borysow}, A., {Frommhold}, L., \& {Moraldi}, M. 1989, \apj, 336, 495, \dodoi{10.1086/167027}

\bibitem[{{Borysow} {et~al.}(2001){Borysow}, {Jorgensen}, \& {Fu}}]{2001JQSRT..68..235B}
{Borysow}, A., {Jorgensen}, U.~G., \& {Fu}, Y. 2001, \jqsrt, 68, 235, \dodoi{10.1016/S0022-4073(00)00023-6}

\bibitem[{{Borysow} {et~al.}(1988){Borysow}, {Frommhold}, \& {Birnbaum}}]{1988ApJ...326..509B}
{Borysow}, J., {Frommhold}, L., \& {Birnbaum}, G. 1988, \apj, 326, 509, \dodoi{10.1086/166112}

\bibitem[{{Brewer} {et~al.}(2017){Brewer}, {Fischer}, \& {Madhusudhan}}]{2017AJ....153...83B}
{Brewer}, J.~M., {Fischer}, D.~A., \& {Madhusudhan}, N. 2017, \aj, 153, 83, \dodoi{10.3847/1538-3881/153/2/83}

\bibitem[{{Buchner} {et~al.}(2014){Buchner}, {Georgakakis}, {Nandra}, {Hsu}, {Rangel}, {Brightman}, {Merloni}, {Salvato}, {Donley}, \& {Kocevski}}]{2014A&A...564A.125B}
{Buchner}, J., {Georgakakis}, A., {Nandra}, K., {et~al.} 2014, \aap, 564, A125, \dodoi{10.1051/0004-6361/201322971}

\bibitem[{{Bushouse} {et~al.}(2022){Bushouse}, {Eisenhamer}, {Dencheva}, {Davies}, {Greenfield}, {Morrison}, {Hodge}, {Simon}, {Grumm}, {Droettboom}, {Slavich}, {Sosey}, {Pauly}, {Miller}, {Jedrzejewski}, {Hack}, {Davis}, {Crawford}, {Law}, {Gordon}, {Regan}, {Cara}, {MacDonald}, {Bradley}, {Shanahan}, {Jamieson}, {Teodoro}, \& {Williams}}]{2022zndo...7071140B}
{Bushouse}, H., {Eisenhamer}, J., {Dencheva}, N., {et~al.} 2022, {JWST Calibration Pipeline}, 1.7.0,  Zenodo, \dodoi{10.5281/zenodo.7071140}

\bibitem[{{Chan} \& {Dalgarno}(1965)}]{1965PPS....85..227C}
{Chan}, Y.~M., \& {Dalgarno}, A. 1965, Proceedings of the Physical Society, 85, 227, \dodoi{10.1088/0370-1328/85/2/304}

\bibitem[{{Changeat} {et~al.}(2022){Changeat}, {Edwards}, {Al-Refaie}, {Tsiaras}, {Skinner}, {Cho}, {Yip}, {Anisman}, {Ikoma}, {Bieger}, {Venot}, {Shibata}, {Waldmann}, \& {Tinetti}}]{2022ApJS..260....3C}
{Changeat}, Q., {Edwards}, B., {Al-Refaie}, A.~F., {et~al.} 2022, \apjs, 260, 3, \dodoi{10.3847/1538-4365/ac5cc2}

\bibitem[{{Cort{\'e}s-Zuleta} {et~al.}(2020){Cort{\'e}s-Zuleta}, {Rojo}, {Wang}, {Hinse}, {Hoyer}, {Sanhueza}, {Correa-Amaro}, \& {Albornoz}}]{2020A&A...636A..98C}
{Cort{\'e}s-Zuleta}, P., {Rojo}, P., {Wang}, S., {et~al.} 2020, \aap, 636, A98, \dodoi{10.1051/0004-6361/201936279}

\bibitem[{{Costa} {et~al.}(2017){Costa}, {Jacobson}, \& {Fegley}}]{2017Icar..289...42C}
{Costa}, G. C.~C., {Jacobson}, N.~S., \& {Fegley}, Jr., B. 2017, \icarus, 289, 42, \dodoi{10.1016/j.icarus.2017.02.006}

\bibitem[{{Coulombe} {et~al.}(2023){Coulombe}, {Benneke}, {Challener}, {Piette}, {Wiser}, {Mansfield}, {MacDonald}, {Beltz}, {Feinstein}, {Radica}, {Savel}, {Dos Santos}, {Bean}, {Parmentier}, {Wong}, {Rauscher}, {Komacek}, {Kempton}, {Tan}, {Hammond}, {Lewis}, {Line}, {Lee}, {Shivkumar}, {Crossfield}, {Nixon}, {Rackham}, {Wakeford}, {Welbanks}, {Zhang}, {Batalha}, {Berta-Thompson}, {Changeat}, {D{\'e}sert}, {Espinoza}, {Goyal}, {Harrington}, {Knutson}, {Kreidberg}, {L{\'o}pez-Morales}, {Shporer}, {Sing}, {Stevenson}, {Aggarwal}, {Ahrer}, {Alam}, {Bell}, {Blecic}, {Caceres}, {Carter}, {Casewell}, {Crouzet}, {Cubillos}, {Decin}, {Fortney}, {Gibson}, {Heng}, {Henning}, {Iro}, {Kendrew}, {Lagage}, {Leconte}, {Lendl}, {Lothringer}, {Mancini}, {Mikal-Evans}, {Molaverdikhani}, {Nikolov}, {Ohno}, {Palle}, {Piaulet}, {Redfield}, {Roy}, {Tsai}, {Venot}, \& {Wheatley}}]{2023Natur.620..292C}
{Coulombe}, L.-P., {Benneke}, B., {Challener}, R., {et~al.} 2023, \nat, 620, 292, \dodoi{10.1038/s41586-023-06230-1}

\bibitem[{{Coulombe} {et~al.}(2025){Coulombe}, {Radica}, {Benneke}, {D'Aoust}, {Dang}, {Cowan}, {Parmentier}, {Albert}, {Lafreni{\`e}re}, {Taylor}, {Roy}, {Pelletier}, {Allart}, {Artigau}, {Doyon}, {Jayawardhana}, {Johnstone}, {Kaltenegger}, {Langeveld}, {MacDonald}, {Rowe}, \& {Turner}}]{2025NatAs.tmp...56C}
{Coulombe}, L.-P., {Radica}, M., {Benneke}, B., {et~al.} 2025, Nature Astronomy, \dodoi{10.1038/s41550-025-02488-9}

\bibitem[{{Cridland} {et~al.}(2019){Cridland}, {van Dishoeck}, {Alessi}, \& {Pudritz}}]{2019A&A...632A..63C}
{Cridland}, A.~J., {van Dishoeck}, E.~F., {Alessi}, M., \& {Pudritz}, R.~E. 2019, \aap, 632, A63, \dodoi{10.1051/0004-6361/201936105}

\bibitem[{{Dalgarno} \& {Williams}(1962)}]{1962ApJ...136..690D}
{Dalgarno}, A., \& {Williams}, D.~A. 1962, \apj, 136, 690, \dodoi{10.1086/147428}

\bibitem[{{Deka} {et~al.}(2025){Deka}, {Majumdar}, {Basra Khan}, {Dewan}, {Ghosh}, {Das}, \& {Patra}}]{2025arXiv251027223D}
{Deka}, T., {Majumdar}, L., {Basra Khan}, T., {et~al.} 2025, arXiv e-prints, arXiv:2510.27223, \dodoi{10.48550/arXiv.2510.27223}

\bibitem[{{Edwards} \& {Changeat}(2024)}]{2024ApJ...962L..30E}
{Edwards}, B., \& {Changeat}, Q. 2024, \apjl, 962, L30, \dodoi{10.3847/2041-8213/ad2000}

\bibitem[{{Edwards} {et~al.}(2023){Edwards}, {Changeat}, {Tsiaras}, {Yip}, {Al-Refaie}, {Anisman}, {Bieger}, {Gressier}, {Shibata}, {Skaf}, {Bouwman}, {Cho}, {Ikoma}, {Venot}, {Waldmann}, {Lagage}, \& {Tinetti}}]{2023ApJS..269...31E}
{Edwards}, B., {Changeat}, Q., {Tsiaras}, A., {et~al.} 2023, \apjs, 269, 31, \dodoi{10.3847/1538-4365/ac9f1a}

\bibitem[{{Eistrup} {et~al.}(2018){Eistrup}, {Walsh}, \& {van Dishoeck}}]{2018A&A...613A..14E}
{Eistrup}, C., {Walsh}, C., \& {van Dishoeck}, E.~F. 2018, \aap, 613, A14, \dodoi{10.1051/0004-6361/201731302}

\bibitem[{{Espinoza} {et~al.}(2019){Espinoza}, {Rackham}, {Jord{\'a}n}, {Apai}, {L{\'o}pez-Morales}, {Osip}, {Grimm}, {Hoeijmakers}, {Wilson}, {Bixel}, {McGruder}, {Rodler}, {Weaver}, {Lewis}, {Fortney}, \& {Fraine}}]{2019MNRAS.482.2065E}
{Espinoza}, N., {Rackham}, B.~V., {Jord{\'a}n}, A., {et~al.} 2019, \mnras, 482, 2065, \dodoi{10.1093/mnras/sty2691}

\bibitem[{{Espinoza} {et~al.}(2023){Espinoza}, {{\'U}beda}, {Birkmann}, {Ferruit}, {Valenti}, {Sing}, {Rustamkulov}, {Regan}, {Kendrew}, {Sabbi}, {Schlawin}, {Beatty}, {Albert}, {Greene}, {Nikolov}, {Karakla}, {Keyes}, {Alves de Oliveira}, {B{\"o}ker}, {Pena-Guerrero}, {Giardino}, {Kumari}, {Manjavacas}, {Proffitt}, \& {Rawle}}]{2023PASP..135a8002E}
{Espinoza}, N., {{\'U}beda}, L., {Birkmann}, S.~M., {et~al.} 2023, \pasp, 135, 018002, \dodoi{10.1088/1538-3873/aca3d3}

\bibitem[{{Feinstein} {et~al.}(2023){Feinstein}, {Radica}, {Welbanks}, {Murray}, {Ohno}, {Coulombe}, {Espinoza}, {Bean}, {Teske}, {Benneke}, {Line}, {Rustamkulov}, {Saba}, {Tsiaras}, {Barstow}, {Fortney}, {Gao}, {Knutson}, {MacDonald}, {Mikal-Evans}, {Rackham}, {Taylor}, {Parmentier}, {Batalha}, {Berta-Thompson}, {Carter}, {Changeat}, {dos Santos}, {Gibson}, {Goyal}, {Kreidberg}, {L{\'o}pez-Morales}, {Lothringer}, {Miguel}, {Molaverdikhani}, {Moran}, {Morello}, {Mukherjee}, {Sing}, {Stevenson}, {Wakeford}, {Ahrer}, {Alam}, {Alderson}, {Allen}, {Batalha}, {Bell}, {Blecic}, {Brande}, {Caceres}, {Casewell}, {Chubb}, {Crossfield}, {Crouzet}, {Cubillos}, {Decin}, {D{\'e}sert}, {Harrington}, {Heng}, {Henning}, {Iro}, {Kempton}, {Kendrew}, {Kirk}, {Krick}, {Lagage}, {Lendl}, {Mancini}, {Mansfield}, {May}, {Mayne}, {Nikolov}, {Palle}, {Petit dit de la Roche}, {Piaulet}, {Powell}, {Redfield}, {Rogers}, {Roman}, {Roy}, {Nixon}, {Schlawin}, {Tan}, {Tremblin}, {Turner}, {Venot}, {Waalkes}, {Wheatley}, \&
  {Zhang}}]{2023Natur.614..670F}
{Feinstein}, A.~D., {Radica}, M., {Welbanks}, L., {et~al.} 2023, \nat, 614, 670, \dodoi{10.1038/s41586-022-05674-1}

\bibitem[{{Feroz} {et~al.}(2009){Feroz}, {Hobson}, \& {Bridges}}]{2009MNRAS.398.1601F}
{Feroz}, F., {Hobson}, M.~P., \& {Bridges}, M. 2009, \mnras, 398, 1601, \dodoi{10.1111/j.1365-2966.2009.14548.x}

\bibitem[{{Fonte} {et~al.}(2023){Fonte}, {Turrini}, {Pacetti}, {Schisano}, {Molinari}, {Polychroni}, {Politi}, \& {Changeat}}]{2023MNRAS.520.4683F}
{Fonte}, S., {Turrini}, D., {Pacetti}, E., {et~al.} 2023, \mnras, 520, 4683, \dodoi{10.1093/mnras/stad245}

\bibitem[{{Foreman-Mackey}(2018)}]{celerite2}
{Foreman-Mackey}, D. 2018, Research Notes of the American Astronomical Society, 2, 31, \dodoi{10.3847/2515-5172/aaaf6c}

\bibitem[{{Foreman-Mackey} {et~al.}(2017){Foreman-Mackey}, {Agol}, {Ambikasaran}, \& {Angus}}]{celerite1}
{Foreman-Mackey}, D., {Agol}, E., {Ambikasaran}, S., \& {Angus}, R. 2017, \aj, 154, 220, \dodoi{10.3847/1538-3881/aa9332}

\bibitem[{{Foreman-Mackey} {et~al.}(2013){Foreman-Mackey}, {Hogg}, {Lang}, \& {Goodman}}]{2013PASP..125..306F}
{Foreman-Mackey}, D., {Hogg}, D.~W., {Lang}, D., \& {Goodman}, J. 2013, \pasp, 125, 306, \dodoi{10.1086/670067}

\bibitem[{{Gardner} {et~al.}(2006){Gardner}, {Mather}, {Clampin}, {Doyon}, {Greenhouse}, {Hammel}, {Hutchings}, {Jakobsen}, {Lilly}, {Long}, {Lunine}, {McCaughrean}, {Mountain}, {Nella}, {Rieke}, {Rieke}, {Rix}, {Smith}, {Sonneborn}, {Stiavelli}, {Stockman}, {Windhorst}, \& {Wright}}]{2006SSRv..123..485G}
{Gardner}, J.~P., {Mather}, J.~C., {Clampin}, M., {et~al.} 2006, \ssr, 123, 485, \dodoi{10.1007/s11214-006-8315-7}

\bibitem[{{Gray}(2008)}]{2008oasp.book.....G}
{Gray}, D.~F. 2008, {The Observation and Analysis of Stellar Photospheres} (Cambridge University Press)

\bibitem[{{Gressier} {et~al.}(2025){Gressier}, {Batalha}, {Wogan}, {Alderson}, {Doud}, {Espinoza}, {MacDonald}, {Wakeford}, {Valenti}, {Lewis}, {Seager}, {Stevenson}, {Allen}, {Ca{\~n}as}, {Challener}, {Glidden}, {Huang}, {Lin}, {Louie}, {Maguire}, {Mullens}, {Sotzen}, {Valentine}, {Clampin}, {Pueyo}, {van der Marel}, \& {Mountain}}]{2025AJ....170..292G}
{Gressier}, A., {Batalha}, N.~E., {Wogan}, N., {et~al.} 2025, \aj, 170, 292, \dodoi{10.3847/1538-3881/ae0929}

\bibitem[{{Guillot}(2010)}]{2010A&A...520A..27G}
{Guillot}, T. 2010, \aap, 520, A27, \dodoi{10.1051/0004-6361/200913396}

\bibitem[{{Hargreaves} {et~al.}(2020){Hargreaves}, {Gordon}, {Rey}, {Nikitin}, {Tyuterev}, {Kochanov}, \& {Rothman}}]{2020ApJS..247...55H}
{Hargreaves}, R.~J., {Gordon}, I.~E., {Rey}, M., {et~al.} 2020, \apjs, 247, 55, \dodoi{10.3847/1538-4365/ab7a1a}

\bibitem[{{Harris} {et~al.}(2006){Harris}, {Tennyson}, {Kaminsky}, {Pavlenko}, \& {Jones}}]{2006MNRAS.367..400H}
{Harris}, G.~J., {Tennyson}, J., {Kaminsky}, B.~M., {Pavlenko}, Y.~V., \& {Jones}, H.~R.~A. 2006, \mnras, 367, 400, \dodoi{10.1111/j.1365-2966.2005.09960.x}

\bibitem[{{Hebb} {et~al.}(2010){Hebb}, {Collier-Cameron}, {Triaud}, {Lister}, {Smalley}, {Maxted}, {Hellier}, {Anderson}, {Pollacco}, {Gillon}, {Queloz}, {West}, {Bentley}, {Enoch}, {Haswell}, {Horne}, {Mayor}, {Pepe}, {Segransan}, {Skillen}, {Udry}, \& {Wheatley}}]{2010ApJ...708..224H}
{Hebb}, L., {Collier-Cameron}, A., {Triaud}, A.~H.~M.~J., {et~al.} 2010, \apj, 708, 224, \dodoi{10.1088/0004-637X/708/1/224}

\bibitem[{{Henning} \& {Mutschke}(1997)}]{1997A&A...327..743H}
{Henning}, T., \& {Mutschke}, H. 1997, \aap, 327, 743

\bibitem[{{Huitson} {et~al.}(2013){Huitson}, {Sing}, {Pont}, {Fortney}, {Burrows}, {Wilson}, {Ballester}, {Nikolov}, {Gibson}, {Deming}, {Aigrain}, {Evans}, {Henry}, {Lecavelier des Etangs}, {Showman}, {Vidal-Madjar}, \& {Zahnle}}]{2013MNRAS.434.3252H}
{Huitson}, C.~M., {Sing}, D.~K., {Pont}, F., {et~al.} 2013, \mnras, 434, 3252, \dodoi{10.1093/mnras/stt1243}

\bibitem[{{Inglis} {et~al.}(2024){Inglis}, {Batalha}, {Lewis}, {Kataria}, {Knutson}, {Kilpatrick}, {Gagnebin}, {Mukherjee}, {Pettyjohn}, {Crossfield}, {Foote}, {Grant}, {Henry}, {Lally}, {McKemmish}, {Sing}, {Wakeford}, {Zapata Trujillo}, \& {Zellem}}]{2024ApJ...973L..41I}
{Inglis}, J., {Batalha}, N.~E., {Lewis}, N.~K., {et~al.} 2024, \apjl, 973, L41, \dodoi{10.3847/2041-8213/ad725e}

\bibitem[{{Jaeger} {et~al.}(1994){Jaeger}, {Mutschke}, {Begemann}, {Dorschner}, \& {Henning}}]{1994A&A...292..641J}
{Jaeger}, C., {Mutschke}, H., {Begemann}, B., {Dorschner}, J., \& {Henning}, T. 1994, \aap, 292, 641

\bibitem[{{Jakobsen} {et~al.}(2022){Jakobsen}, {Ferruit}, {Alves de Oliveira}, {Arribas}, {Bagnasco}, {Barho}, {Beck}, {Birkmann}, {B{\"o}ker}, {Bunker}, {Charlot}, {de Jong}, {de Marchi}, {Ehrenwinkler}, {Falcolini}, {Fels}, {Franx}, {Franz}, {Funke}, {Giardino}, {Gnata}, {Holota}, {Honnen}, {Jensen}, {Jentsch}, {Johnson}, {Jollet}, {Karl}, {Kling}, {K{\"o}hler}, {Kolm}, {Kumari}, {Lander}, {Lemke}, {L{\'o}pez-Caniego}, {L{\"u}tzgendorf}, {Maiolino}, {Manjavacas}, {Marston}, {Maschmann}, {Maurer}, {Messerschmidt}, {Moseley}, {Mosner}, {Mott}, {Muzerolle}, {Pirzkal}, {Pittet}, {Plitzke}, {Posselt}, {Rapp}, {Rauscher}, {Rawle}, {Rix}, {R{\"o}del}, {Rumler}, {Sabbi}, {Salvignol}, {Schmid}, {Sirianni}, {Smith}, {Strada}, {te Plate}, {Valenti}, {Wettemann}, {Wiehe}, {Wiesmayer}, {Willott}, {Wright}, {Zeidler}, \& {Zincke}}]{2022A&A...661A..80J}
{Jakobsen}, P., {Ferruit}, P., {Alves de Oliveira}, C., {et~al.} 2022, \aap, 661, A80, \dodoi{10.1051/0004-6361/202142663}

\bibitem[{Kass \& Raftery(1995)}]{Kass01061995}
Kass, R.~E., \& Raftery, A.~E. 1995, Journal of the American Statistical Association, 90, 773, \dodoi{10.1080/01621459.1995.10476572}

\bibitem[{{Khorshid} {et~al.}(2024){Khorshid}, {Min}, {Polman}, \& {Waters}}]{2024A&A...685A..64K}
{Khorshid}, N., {Min}, M., {Polman}, J., \& {Waters}, L.~B.~F.~M. 2024, \aap, 685, A64, \dodoi{10.1051/0004-6361/202347124}

\bibitem[{{Kirk} {et~al.}(2024){Kirk}, {Ahrer}, {Claringbold}, {Zamyatina}, {Fisher}, {McCormack}, {Panwar}, {Powell}, {Taylor}, {Thorngren}, {Christie}, {Esparza-Borges}, {Tsai}, {Alderson}, {Booth}, {Fairman}, {L{\'o}pez-Morales}, {Mayne}, {Meech}, {Molliere}, {Owen}, {Penzlin}, {Sergeev}, {Valentine}, {Wakeford}, \& {Wheatley}}]{2024arXiv241008116K}
{Kirk}, J., {Ahrer}, E.-M., {Claringbold}, A.~B., {et~al.} 2024, arXiv e-prints, arXiv:2410.08116, \dodoi{10.48550/arXiv.2410.08116}

\bibitem[{{Kitzmann} \& {Heng}(2018)}]{2018MNRAS.475...94K}
{Kitzmann}, D., \& {Heng}, K. 2018, \mnras, 475, 94, \dodoi{10.1093/mnras/stx3141}

\bibitem[{{Koike} {et~al.}(1995){Koike}, {Kaito}, {Yamamoto}, {Shibai}, {Kimura}, \& {Suto}}]{1995Icar..114..203K}
{Koike}, C., {Kaito}, C., {Yamamoto}, T., {et~al.} 1995, \icarus, 114, 203, \dodoi{10.1006/icar.1995.1055}

\bibitem[{{Kreidberg}(2015)}]{2015PASP..127.1161K}
{Kreidberg}, L. 2015, \pasp, 127, 1161, \dodoi{10.1086/683602}

\bibitem[{{Louie} {et~al.}(2024){Louie}, {Mullens}, {Alderson}, {Glidden}, {Lewis}, {Wakeford}, {Batalha}, {Col{\'o}n}, {Gressier}, {Long}, {Radica}, {Espinoza}, {Goyal}, {MacDonald}, {May}, {Seager}, {Stevenson}, {Valenti}, {Allen}, {Ca{\~n}as}, {Challener}, {Grant}, {Huang}, {Lin}, {Valentine}, {Perrin}, {Pueyo}, \& {van der Marel}}]{2024arXiv241203675L}
{Louie}, D.~R., {Mullens}, E., {Alderson}, L., {et~al.} 2024, arXiv e-prints, arXiv:2412.03675, \dodoi{10.48550/arXiv.2412.03675}

\bibitem[{{Lustig-Yaeger} {et~al.}(2023){Lustig-Yaeger}, {Fu}, {May}, {Ceballos}, {Moran}, {Peacock}, {Stevenson}, {Kirk}, {L{\'o}pez-Morales}, {MacDonald}, {Mayorga}, {Sing}, {Sotzen}, {Valenti}, {Redai}, {Alam}, {Batalha}, {Bennett}, {Gonzalez-Quiles}, {Kruse}, {Lothringer}, {Rustamkulov}, \& {Wakeford}}]{2023NatAs...7.1317L}
{Lustig-Yaeger}, J., {Fu}, G., {May}, E.~M., {et~al.} 2023, Nature Astronomy, 7, 1317, \dodoi{10.1038/s41550-023-02064-z}

\bibitem[{{Madhusudhan} {et~al.}(2014){Madhusudhan}, {Amin}, \& {Kennedy}}]{2014ApJ...794L..12M}
{Madhusudhan}, N., {Amin}, M.~A., \& {Kennedy}, G.~M. 2014, \apjl, 794, L12, \dodoi{10.1088/2041-8205/794/1/L12}

\bibitem[{{Madhusudhan} {et~al.}(2011){Madhusudhan}, {Harrington}, {Stevenson}, {Nymeyer}, {Campo}, {Wheatley}, {Deming}, {Blecic}, {Hardy}, {Lust}, {Anderson}, {Collier-Cameron}, {Britt}, {Bowman}, {Hebb}, {Hellier}, {Maxted}, {Pollacco}, \& {West}}]{2011Natur.469...64M}
{Madhusudhan}, N., {Harrington}, J., {Stevenson}, K.~B., {et~al.} 2011, \nat, 469, 64, \dodoi{10.1038/nature09602}

\bibitem[{{Mandell} {et~al.}(2013){Mandell}, {Haynes}, {Sinukoff}, {Madhusudhan}, {Burrows}, \& {Deming}}]{2013ApJ...779..128M}
{Mandell}, A.~M., {Haynes}, K., {Sinukoff}, E., {et~al.} 2013, \apj, 779, 128, \dodoi{10.1088/0004-637X/779/2/128}

\bibitem[{{Mansfield} {et~al.}(2018){Mansfield}, {Bean}, {Line}, {Parmentier}, {Kreidberg}, {D{\'e}sert}, {Fortney}, {Stevenson}, {Arcangeli}, \& {Dragomir}}]{2018AJ....156...10M}
{Mansfield}, M., {Bean}, J.~L., {Line}, M.~R., {et~al.} 2018, \aj, 156, 10, \dodoi{10.3847/1538-3881/aac497}

\bibitem[{{May} {et~al.}(2023){May}, {MacDonald}, {Bennett}, {Moran}, {Wakeford}, {Peacock}, {Lustig-Yaeger}, {Highland}, {Stevenson}, {Sing}, {Mayorga}, {Batalha}, {Kirk}, {L{\'o}pez-Morales}, {Valenti}, {Alam}, {Alderson}, {Fu}, {Gonzalez-Quiles}, {Lothringer}, {Rustamkulov}, \& {Sotzen}}]{2023ApJ...959L...9M}
{May}, E.~M., {MacDonald}, R.~J., {Bennett}, K.~A., {et~al.} 2023, \apjl, 959, L9, \dodoi{10.3847/2041-8213/ad054f}

\bibitem[{{Mayo} {et~al.}(2025){Mayo}, {Fortenbach}, {Louie}, {Dressing}, {Turtelboom}, {Giacalone}, \& {Harada}}]{2025AJ....170...50M}
{Mayo}, A.~W., {Fortenbach}, C.~D., {Louie}, D.~R., {et~al.} 2025, \aj, 170, 50, \dodoi{10.3847/1538-3881/adda2e}

\bibitem[{{Meech} {et~al.}(2025){Meech}, {Claringbold}, {Ahrer}, {Kirk}, {L{\'o}pez-Morales}, {Taylor}, {Booth}, {Penzlin}, {Alderson}, {Christie}, {Esparza-Borges}, {Fairman}, {Mayne}, {McCormack}, {Owen}, {Panwar}, {Powell}, {Sergeev}, {Valentine}, {Wakeford}, {Wheatley}, \& {Zamyatina}}]{2025MNRAS.539.1381M}
{Meech}, A., {Claringbold}, A.~B., {Ahrer}, E.-M., {et~al.} 2025, \mnras, 539, 1381, \dodoi{10.1093/mnras/staf530}

\bibitem[{{Molli{\`e}re} {et~al.}(2019){Molli{\`e}re}, {Wardenier}, {van Boekel}, {Henning}, {Molaverdikhani}, \& {Snellen}}]{2019A&A...627A..67M}
{Molli{\`e}re}, P., {Wardenier}, J.~P., {van Boekel}, R., {et~al.} 2019, \aap, 627, A67, \dodoi{10.1051/0004-6361/201935470}

\bibitem[{{Mordasini} {et~al.}(2016){Mordasini}, {van Boekel}, {Molli{\`e}re}, {Henning}, \& {Benneke}}]{2016ApJ...832...41M}
{Mordasini}, C., {van Boekel}, R., {Molli{\`e}re}, P., {Henning}, T., \& {Benneke}, B. 2016, \apj, 832, 41, \dodoi{10.3847/0004-637X/832/1/41}

\bibitem[{{Nasedkin} {et~al.}(2024){Nasedkin}, {Molli{\`e}re}, \& {Blain}}]{2024JOSS....9.5875N}
{Nasedkin}, E., {Molli{\`e}re}, P., \& {Blain}, D. 2024, The Journal of Open Source Software, 9, 5875, \dodoi{10.21105/joss.05875}

\bibitem[{{{\"O}berg} {et~al.}(2011){{\"O}berg}, {Murray-Clay}, \& {Bergin}}]{2011ApJ...743L..16O}
{{\"O}berg}, K.~I., {Murray-Clay}, R., \& {Bergin}, E.~A. 2011, \apjl, 743, L16, \dodoi{10.1088/2041-8205/743/1/L16}

\bibitem[{{Petz} {et~al.}(2024){Petz}, {Johnson}, {Asnodkar}, {Wang}, {Gaudi}, {Henning}, {Keles}, {Molaverdikhani}, {Poppenhaeger}, {Scandariato}, {Shkolnik}, {Sicilia}, {Strassmeier}, \& {Yan}}]{2024MNRAS.527.7079P}
{Petz}, S., {Johnson}, M.~C., {Asnodkar}, A.~P., {et~al.} 2024, \mnras, 527, 7079, \dodoi{10.1093/mnras/stad3481}

\bibitem[{{Polyansky} {et~al.}(2018){Polyansky}, {Kyuberis}, {Zobov}, {Tennyson}, {Yurchenko}, \& {Lodi}}]{2018MNRAS.480.2597P}
{Polyansky}, O.~L., {Kyuberis}, A.~A., {Zobov}, N.~F., {et~al.} 2018, \mnras, 480, 2597, \dodoi{10.1093/mnras/sty1877}

\bibitem[{{Posch} {et~al.}(2003){Posch}, {Kerschbaum}, {Fabian}, {Mutschke}, {Dorschner}, {Tamanai}, \& {Henning}}]{2003ApJS..149..437P}
{Posch}, T., {Kerschbaum}, F., {Fabian}, D., {et~al.} 2003, \apjs, 149, 437, \dodoi{10.1086/379167}

\bibitem[{{Powell} {et~al.}(2024){Powell}, {Feinstein}, {Lee}, {Zhang}, {Tsai}, {Taylor}, {Kirk}, {Bell}, {Barstow}, {Gao}, {Bean}, {Blecic}, {Chubb}, {Crossfield}, {Jordan}, {Kitzmann}, {Moran}, {Morello}, {Moses}, {Welbanks}, {Yang}, {Zhang}, {Ahrer}, {Bello-Arufe}, {Brande}, {Casewell}, {Crouzet}, {Cubillos}, {Demory}, {Dyrek}, {Flagg}, {Hu}, {Inglis}, {Jones}, {Kreidberg}, {L{\'o}pez-Morales}, {Lagage}, {Meier Vald{\'e}s}, {Miguel}, {Parmentier}, {Piette}, {Rackham}, {Radica}, {Redfield}, {Stevenson}, {Wakeford}, {Aggarwal}, {Alam}, {Batalha}, {Batalha}, {Benneke}, {Berta-Thompson}, {Brady}, {Caceres}, {Carter}, {D{\'e}sert}, {Harrington}, {Iro}, {Line}, {Lothringer}, {MacDonald}, {Mancini}, {Molaverdikhani}, {Mukherjee}, {Nixon}, {Oza}, {Palle}, {Rustamkulov}, {Sing}, {Steinrueck}, {Venot}, {Wheatley}, \& {Yurchenko}}]{2024Natur.626..979P}
{Powell}, D., {Feinstein}, A.~D., {Lee}, E. K.~H., {et~al.} 2024, \nat, 626, 979, \dodoi{10.1038/s41586-024-07040-9}

\bibitem[{{Radica}(2024)}]{2024JOSS....9.6898R}
{Radica}, M. 2024, The Journal of Open Source Software, 9, 6898, \dodoi{10.21105/joss.06898}

\bibitem[{{Radica} {et~al.}(2024){Radica}, {Coulombe}, {Taylor}, {Albert}, {Allart}, {Benneke}, {Cowan}, {Dang}, {Lafreni{\`e}re}, {Thorngren}, {Artigau}, {Doyon}, {Flagg}, {Johnstone}, {Pelletier}, \& {Roy}}]{2024ApJ...962L..20R}
{Radica}, M., {Coulombe}, L.-P., {Taylor}, J., {et~al.} 2024, \apjl, 962, L20, \dodoi{10.3847/2041-8213/ad20e4}

\bibitem[{{Ridden-Harper} {et~al.}(2023){Ridden-Harper}, {de Mooij}, {Jayawardhana}, {Gibson}, {Karjalainen}, \& {Karjalainen}}]{2023AJ....165..211R}
{Ridden-Harper}, A., {de Mooij}, E., {Jayawardhana}, R., {et~al.} 2023, \aj, 165, 211, \dodoi{10.3847/1538-3881/acc654}

\bibitem[{{Rothman} {et~al.}(2010){Rothman}, {Gordon}, {Barber}, {Dothe}, {Gamache}, {Goldman}, {Perevalov}, {Tashkun}, \& {Tennyson}}]{2010JQSRT.111.2139R}
{Rothman}, L.~S., {Gordon}, I.~E., {Barber}, R.~J., {et~al.} 2010, \jqsrt, 111, 2139, \dodoi{10.1016/j.jqsrt.2010.05.001}

\bibitem[{{Rothman} {et~al.}(2013){Rothman}, {Gordon}, {Babikov}, {Barbe}, {Chris Benner}, {Bernath}, {Birk}, {Bizzocchi}, {Boudon}, {Brown}, {Campargue}, {Chance}, {Cohen}, {Coudert}, {Devi}, {Drouin}, {Fayt}, {Flaud}, {Gamache}, {Harrison}, {Hartmann}, {Hill}, {Hodges}, {Jacquemart}, {Jolly}, {Lamouroux}, {Le Roy}, {Li}, {Long}, {Lyulin}, {Mackie}, {Massie}, {Mikhailenko}, {M{\"u}ller}, {Naumenko}, {Nikitin}, {Orphal}, {Perevalov}, {Perrin}, {Polovtseva}, {Richard}, {Smith}, {Starikova}, {Sung}, {Tashkun}, {Tennyson}, {Toon}, {Tyuterev}, \& {Wagner}}]{2013JQSRT.130....4R}
{Rothman}, L.~S., {Gordon}, I.~E., {Babikov}, Y., {et~al.} 2013, \jqsrt, 130, 4, \dodoi{10.1016/j.jqsrt.2013.07.002}

\bibitem[{{Saha}(2023)}]{2023ApJS..268....2S}
{Saha}, S. 2023, \apjs, 268, 2, \dodoi{10.3847/1538-4365/acdb6b}

\bibitem[{{Saha}(2024)}]{2024ApJS..274...13S}
---. 2024, \apjs, 274, 13, \dodoi{10.3847/1538-4365/ad6a60}

\bibitem[{{Saha}(2025)}]{2025MNRAS.539..928S}
---. 2025, \mnras, 539, 928, \dodoi{10.1093/mnras/staf550}

\bibitem[{{Saha} {et~al.}(2021){Saha}, {Chakrabarty}, \& {Sengupta}}]{2021AJ....162...18S}
{Saha}, S., {Chakrabarty}, A., \& {Sengupta}, S. 2021, \aj, 162, 18, \dodoi{10.3847/1538-3881/ac01dd}

\bibitem[{{Saha} \& {Jenkins}(2025{\natexlab{a}})}]{saha2025a}
{Saha}, S., \& {Jenkins}, J.~S. 2025{\natexlab{a}}, \apjl, 994, L39, \dodoi{10.3847/2041-8213/ae1c1c}

\bibitem[{{Saha} \& {Jenkins}(2025{\natexlab{b}})}]{saha2025b}
---. 2025{\natexlab{b}}, arXiv e-prints, arXiv:2510.11479, \dodoi{10.48550/arXiv.2510.11479}

\bibitem[{{Saha} \& {Sengupta}(2021)}]{2021AJ....162..221S}
{Saha}, S., \& {Sengupta}, S. 2021, \aj, 162, 221, \dodoi{10.3847/1538-3881/ac294d}

\bibitem[{{Saha} {et~al.}(2025){Saha}, {Jenkins}, {Parmentier}, {Hoyer}, {Deleuil}, {Crossfield}, {Pe{\~n}a R.}, {Vines}, {Ram{\'\i}rez Reyes}, \& {D{\'\i}az}}]{2025A&A...700A..45S}
{Saha}, S., {Jenkins}, J.~S., {Parmentier}, V., {et~al.} 2025, \aap, 700, A45, \dodoi{10.1051/0004-6361/202554303}

\bibitem[{{Schneider} \& {Bitsch}(2021)}]{2021A&A...654A..71S}
{Schneider}, A.~D., \& {Bitsch}, B. 2021, \aap, 654, A71, \dodoi{10.1051/0004-6361/202039640}

\bibitem[{{Sedaghati} {et~al.}(2017){Sedaghati}, {Boffin}, {MacDonald}, {Gandhi}, {Madhusudhan}, {Gibson}, {Oshagh}, {Claret}, \& {Rauer}}]{2017Natur.549..238S}
{Sedaghati}, E., {Boffin}, H. M.~J., {MacDonald}, R.~J., {et~al.} 2017, \nat, 549, 238, \dodoi{10.1038/nature23651}

\bibitem[{{Sedaghati} {et~al.}(2021){Sedaghati}, {MacDonald}, {Casasayas-Barris}, {Hoeijmakers}, {Boffin}, {Rodler}, {Brahm}, {Jones}, {S{\'a}nchez-L{\'o}pez}, {Carleo}, {Figueira}, {Mehner}, \& {L{\'o}pez-Puertas}}]{2021MNRAS.505..435S}
{Sedaghati}, E., {MacDonald}, R.~J., {Casasayas-Barris}, N., {et~al.} 2021, \mnras, 505, 435, \dodoi{10.1093/mnras/stab1164}

\bibitem[{{Servoin} \& {Piriou}(1973)}]{1973PSSBR..55..677S}
{Servoin}, J.~L., \& {Piriou}, B. 1973, Physica Status Solidi B Basic Research, 55, 677, \dodoi{10.1002/pssb.2220550224}

\bibitem[{{Shornikov}(2019)}]{2019RJPCA..93.1024S}
{Shornikov}, S.~I. 2019, Russian Journal of Physical Chemistry A, 93, 1024, \dodoi{10.1134/S0036024419060293}

\bibitem[{{Siefke} {et~al.}(2016){Siefke}, {Kroker}, {Pfeiffer}, {Puffky}, {Dietrich}, {Franta}, {Ohl{\'\i}dal}, {Szeghalmi}, {Kley}, \& {T{\"u}nnermann}}]{2016arXiv160704866S}
{Siefke}, T., {Kroker}, S., {Pfeiffer}, K., {et~al.} 2016, arXiv e-prints, arXiv:1607.04866, \dodoi{10.48550/arXiv.1607.04866}

\bibitem[{{Smith} {et~al.}(2024){Smith}, {Sanchez}, {Line}, {Rauscher}, {Mansfield}, {Kempton}, {Savel}, {Wardenier}, {Pino}, {Bean}, {Beltz}, {Panwar}, {Brogi}, {Malsky}, {Fortney}, {D{\'e}sert}, {Pelletier}, {Parmentier}, {Kanumalla}, {Welbanks}, {Meyer}, \& {Monnier}}]{2024AJ....168..293S}
{Smith}, P. C.~B., {Sanchez}, J.~A., {Line}, M.~R., {et~al.} 2024, \aj, 168, 293, \dodoi{10.3847/1538-3881/ad8574}

\bibitem[{{Sousa-Silva} {et~al.}(2015){Sousa-Silva}, {Al-Refaie}, {Tennyson}, \& {Yurchenko}}]{2015MNRAS.446.2337S}
{Sousa-Silva}, C., {Al-Refaie}, A.~F., {Tennyson}, J., \& {Yurchenko}, S.~N. 2015, \mnras, 446, 2337, \dodoi{10.1093/mnras/stu2246}

\bibitem[{{Speagle}(2020)}]{2020MNRAS.493.3132S}
{Speagle}, J.~S. 2020, \mnras, 493, 3132, \dodoi{10.1093/mnras/staa278}

\bibitem[{Stull(1947)}]{Stull1947}
Stull, D.~R. 1947, Industrial \& Engineering Chemistry, 39, 517, \dodoi{10.1021/ie50448a022}

\bibitem[{{Tumborang} {et~al.}(2024){Tumborang}, {Spake}, {Knutson}, {Weiner Mansfield}, {Paragas}, {Edwards}, {Kataria}, {Evans-Soma}, {Lewis}, \& {Ballester}}]{2024AJ....168..296T}
{Tumborang}, A.~A., {Spake}, J.~J., {Knutson}, H.~A., {et~al.} 2024, \aj, 168, 296, \dodoi{10.3847/1538-3881/ad863f}

\bibitem[{{Ueda} {et~al.}(1998){Ueda}, {Yanagi}, {Noshiro}, {Hosono}, \& {Kawazoe}}]{1998JPCM...10.3669U}
{Ueda}, K., {Yanagi}, H., {Noshiro}, R., {Hosono}, H., \& {Kawazoe}, H. 1998, Journal of Physics Condensed Matter, 10, 3669, \dodoi{10.1088/0953-8984/10/16/018}

\bibitem[{{Weiner Mansfield} {et~al.}(2024){Weiner Mansfield}, {Line}, {Wardenier}, {Brogi}, {Bean}, {Beltz}, {Smith}, {Zalesky}, {Batalha}, {Kempton}, {Montet}, {Owen}, {Plavchan}, \& {Rauscher}}]{2024AJ....168...14W}
{Weiner Mansfield}, M., {Line}, M.~R., {Wardenier}, J.~P., {et~al.} 2024, \aj, 168, 14, \dodoi{10.3847/1538-3881/ad4a5f}

\bibitem[{{Wiser} {et~al.}(2025){Wiser}, {Bell}, {Line}, {Schlawin}, {Beatty}, {Welbanks}, {Greene}, {Parmentier}, {Murphy}, {Fortney}, {Arnold}, {Mehta}, {Ohno}, \& {Mukherjee}}]{2025arXiv250601800W}
{Wiser}, L.~S., {Bell}, T.~J., {Line}, M.~R., {et~al.} 2025, arXiv e-prints, arXiv:2506.01800, \dodoi{10.48550/arXiv.2506.01800}

\bibitem[{{Wong} {et~al.}(2016){Wong}, {Knutson}, {Kataria}, {Lewis}, {Burrows}, {Fortney}, {Schwartz}, {Shporer}, {Agol}, {Cowan}, {Deming}, {D{\'e}sert}, {Fulton}, {Howard}, {Langton}, {Laughlin}, {Showman}, \& {Todorov}}]{2016ApJ...823..122W}
{Wong}, I., {Knutson}, H.~A., {Kataria}, T., {et~al.} 2016, \apj, 823, 122, \dodoi{10.3847/0004-637X/823/2/122}

\bibitem[{{Xue} {et~al.}(2024){Xue}, {Bean}, {Zhang}, {Welbanks}, {Lunine}, \& {August}}]{2024ApJ...963L...5X}
{Xue}, Q., {Bean}, J.~L., {Zhang}, M., {et~al.} 2024, \apjl, 963, L5, \dodoi{10.3847/2041-8213/ad2682}

\bibitem[{{Zeidler} {et~al.}(2011){Zeidler}, {Posch}, {Mutschke}, {Richter}, \& {Wehrhan}}]{2011A&A...526A..68Z}
{Zeidler}, S., {Posch}, T., {Mutschke}, H., {Richter}, H., \& {Wehrhan}, O. 2011, \aap, 526, A68, \dodoi{10.1051/0004-6361/201015219}

\end{thebibliography}

\clearpage

\section*{Appendix}

\renewcommand{\thefigure}{A\arabic{figure}}
\renewcommand{\thetable}{A\arabic{table}}
\renewcommand{\theequation}{A\arabic{equation}}
\renewcommand{\thepage}{A\arabic{page}}
\setcounter{figure}{0}
\setcounter{table}{0}
\setcounter{equation}{0}
\setcounter{page}{1}

\begin{figure*}[!h]
\centering
\begin{tabular}{cc}
\includegraphics[width=0.5\columnwidth]{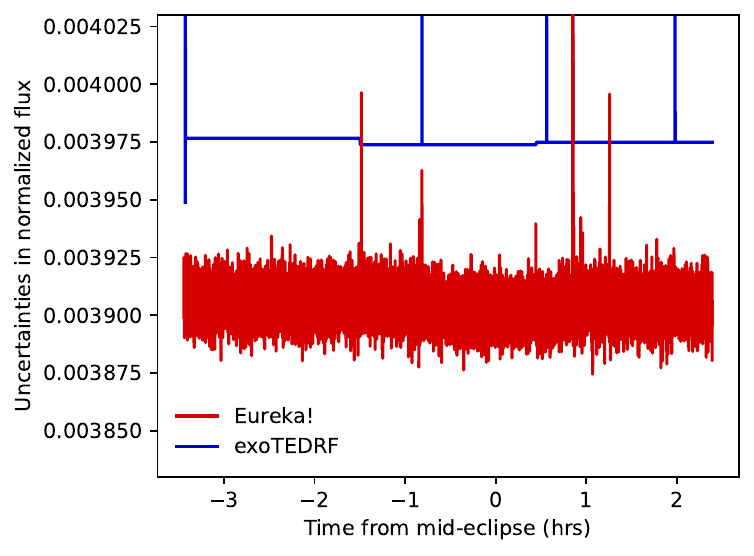} & \includegraphics[width=0.5\columnwidth]{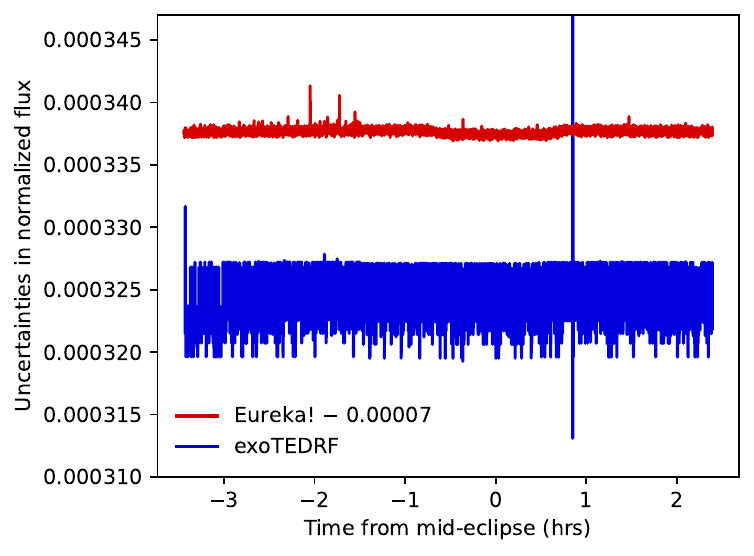}
\end{tabular}
\caption{\textit{Left:} Comparison of uncertainties in the 100th spectroscopic lightcurve from Eureka! and exoTEDRF. \textit{Right:} Same comparison for the white lightcurves. exoTEDRF shows mostly constant uncertainties with occasional large fluctuations. In the white lightcurves, the fluctuations dominate, reducing the difference between the two pipelines. \label{fig:figAerr}}
\end{figure*}

\begin{table*}[!h]
	\centering
	\caption{Estimated parameters from the white lightcurve fit (Eureka!). \label{tab:tr_fit}}
	\begin{tabular}{l c c}
		\hline
		Parameter & Prior & Value \\
		\hline
		$P$ (days) & fixed & $0.788839092$ \\
		$t_0$ (MJD) & $\mathcal{N}(59993.01,0.003)$ & $59993.008198 \pm 8.9 \times 10^{-5}$ \\
		$a/R_*$ & $\mathcal{N}(3.533,0.3)$ & $3.543 \pm 0.045$ \\
		$i$ (deg) & $\mathcal{N}(79.08,0.3)$ & $79.56 \pm 0.34$ \\
		$R_p/R_*$ & fixed & $0.1441$ \\
		$F_p/F_*$ & $\mathcal{U}(0,0.01)$ & $0.001725 \pm 1.5 \times 10^{-5}$ \\
		$e$ & fixed & $0$ \\
		$\omega$ (deg) & fixed & $90.0$ \\
		$a_0$ & $\mathcal{N}(1,0.005)$ & $0.998514 \pm 1.2 \times 10^{-5}$ \\
		$a_1$ & $\mathcal{N}(0,0.0005)$ & $-0.000103 \pm 1.1 \times 10^{-5}$ \\
		$a_2$ & $\mathcal{N}(0,0.0005)$ & $-0.000267 \pm 1.6 \times 10^{-5}$ \\
		$a_3$ & $\mathcal{N}(0,0.0005)$ & $-9 \times 10^{-6} \pm 1.8 \times 10^{-5}$ \\
		$a_4$ & $\mathcal{N}(0,0.0005)$ & $4 \times 10^{-6} \pm 1.3 \times 10^{-5}$ \\
		\hline
	\end{tabular}
\end{table*}

\begin{figure*}
\includegraphics[width=1\columnwidth]{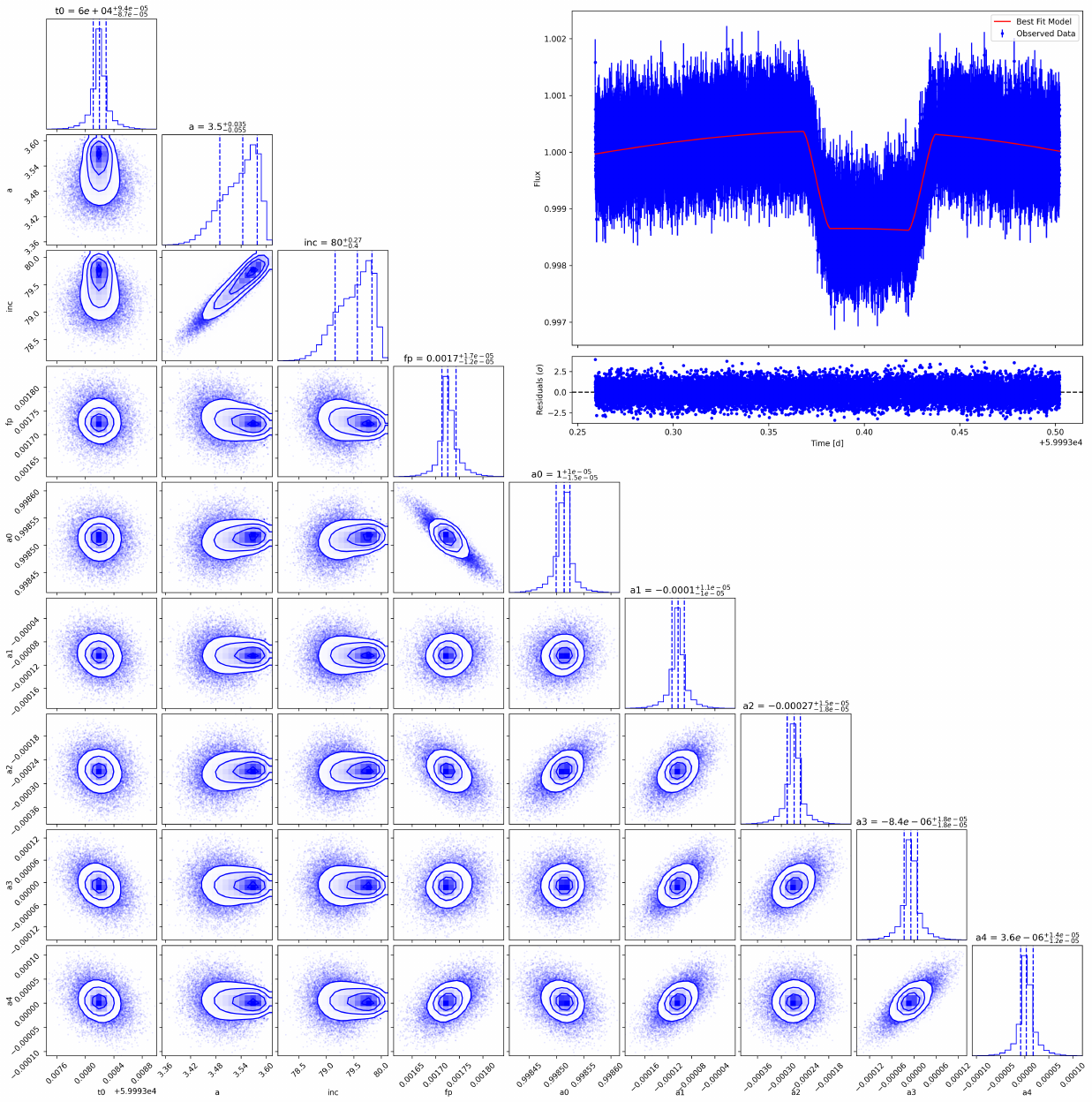}
\caption{The upper-right panel shows the white lightcurve with the best-fit eclipse model, while the bottom panel displays the corner plot of the model's posterior distribution.\label{fig:figA1b}}
\end{figure*}

\begin{figure*}
\centering
\begin{tabular}{cc}
\includegraphics[width=0.5\columnwidth]{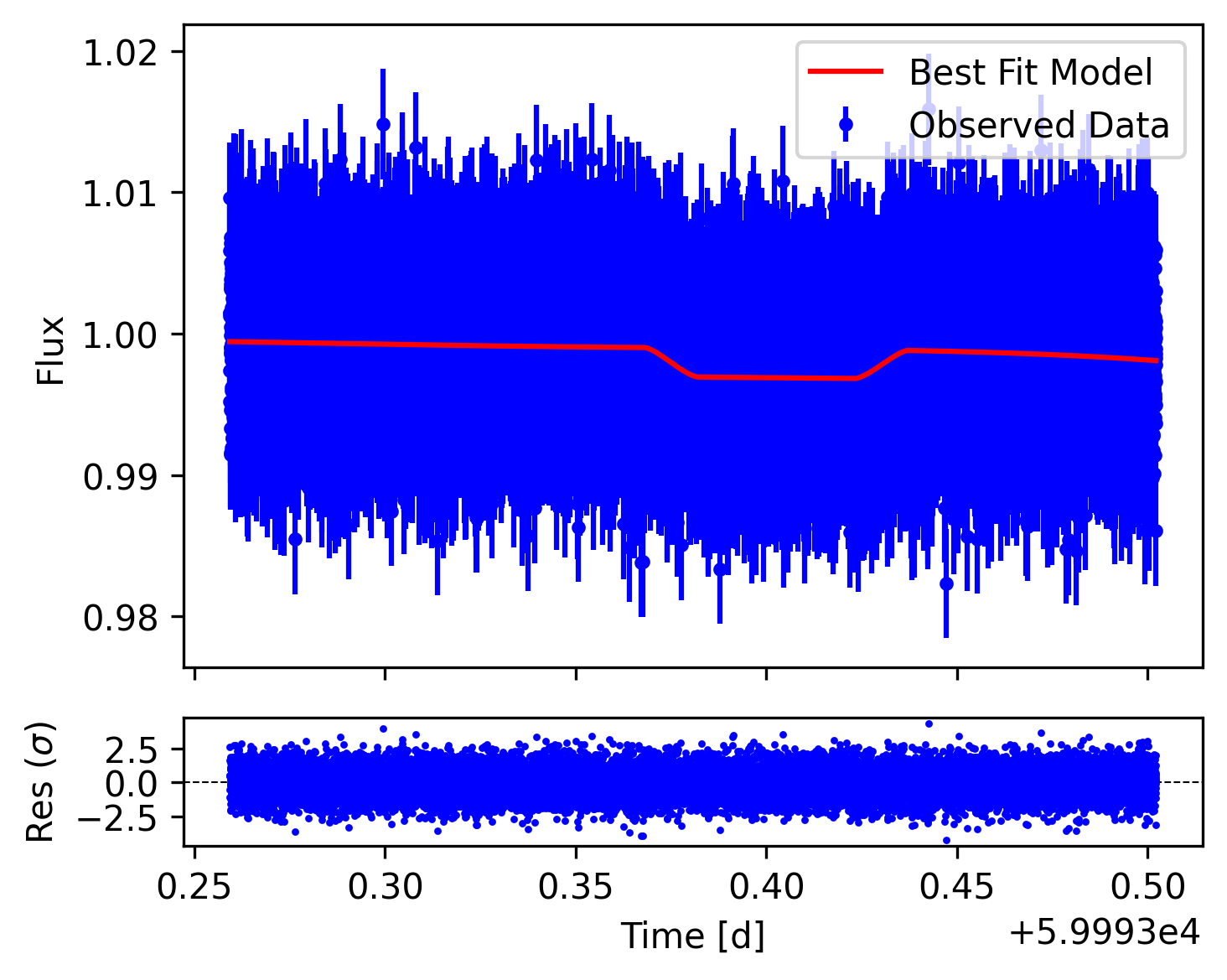} & \includegraphics[width=0.5\columnwidth]{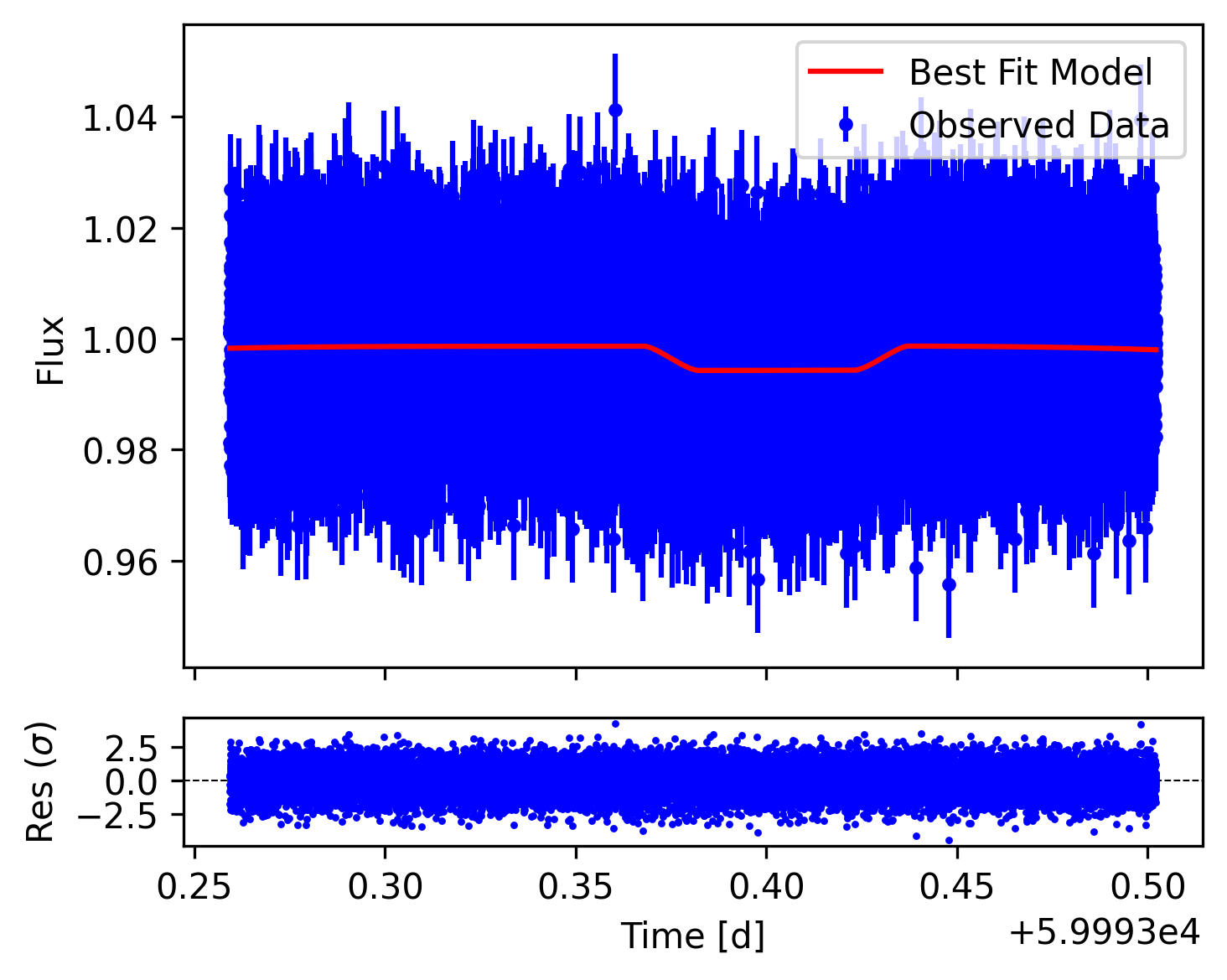}\\
\includegraphics[width=0.5\columnwidth]{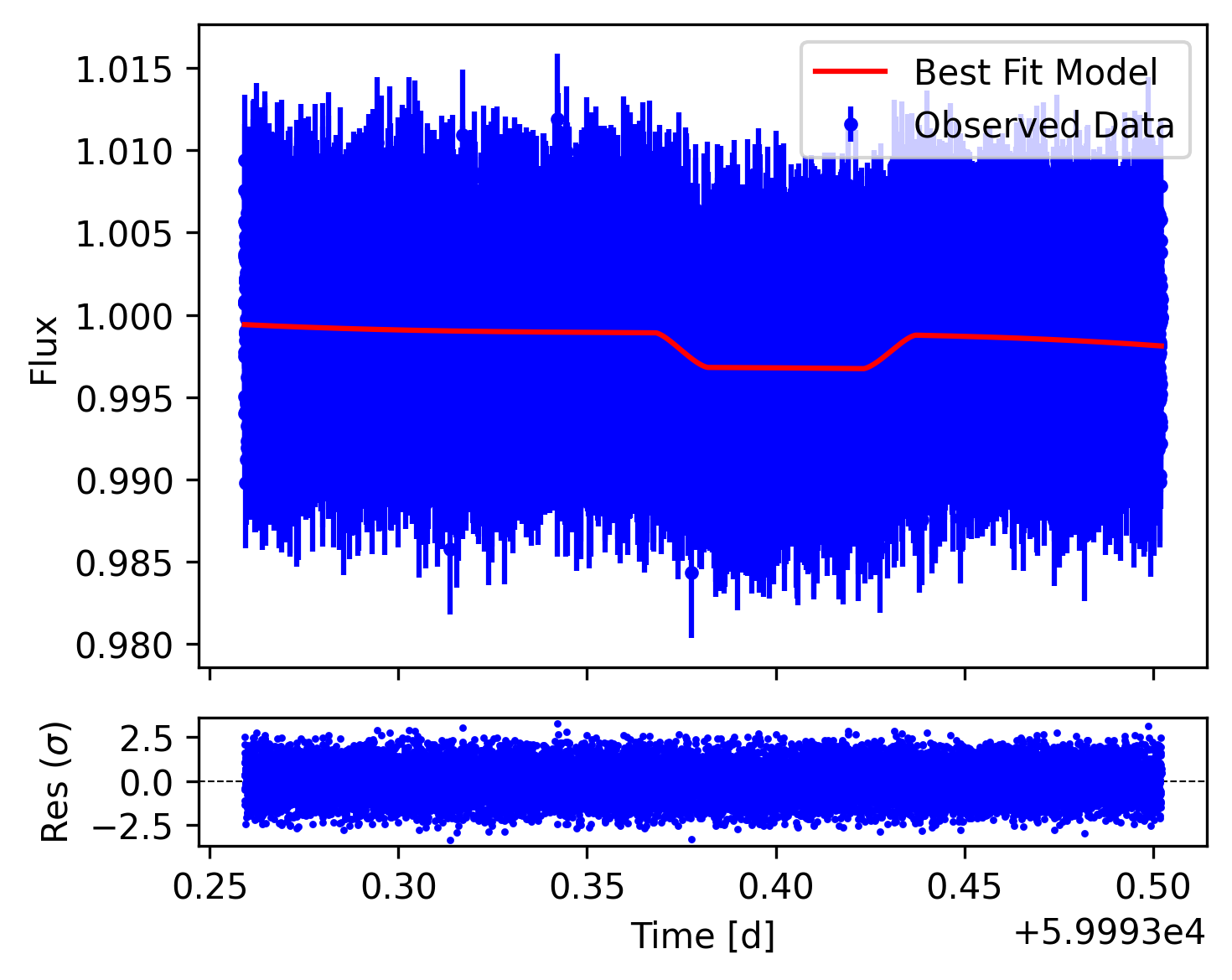} & \includegraphics[width=0.5\columnwidth]{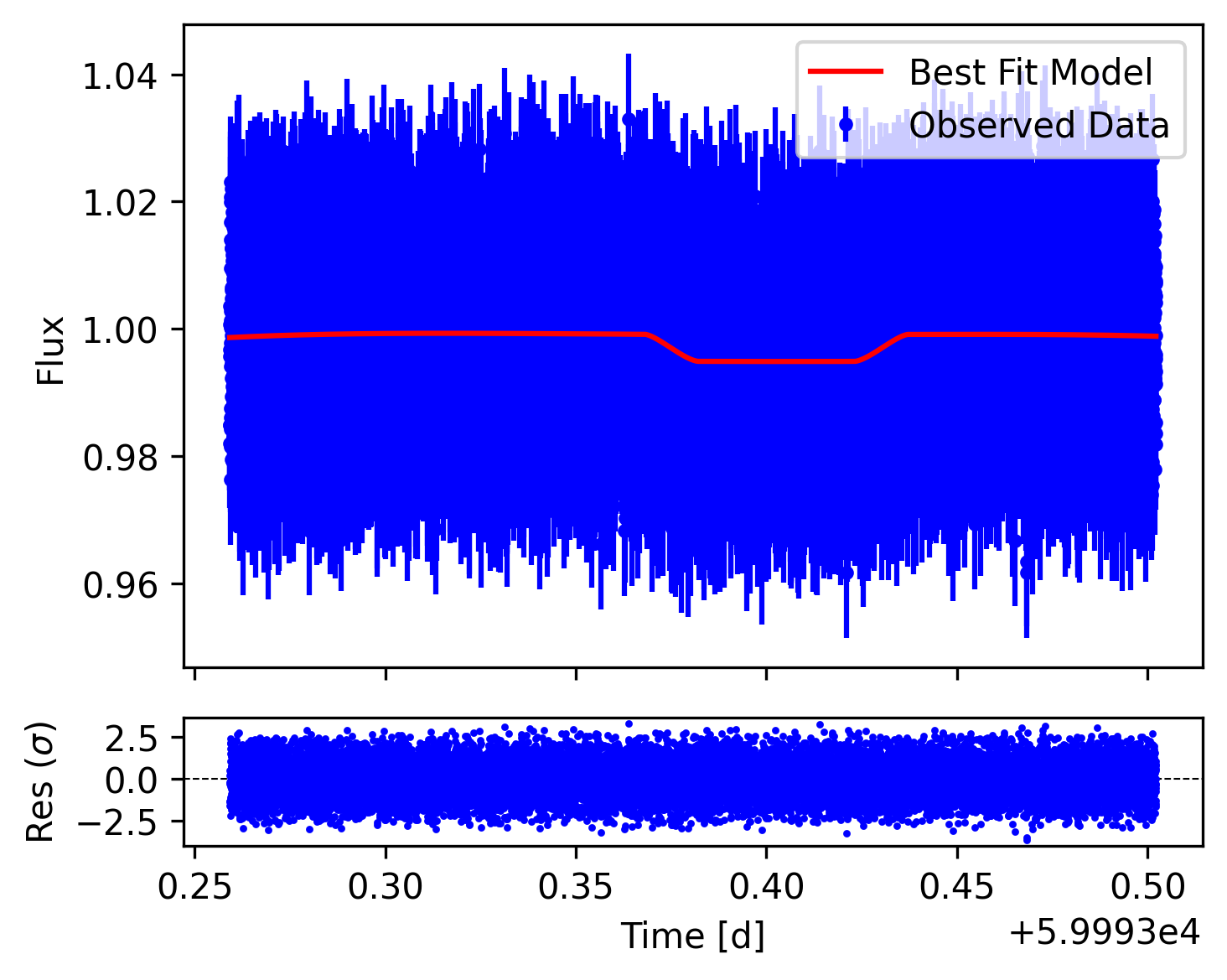}
\end{tabular}
\caption{The top two panels show the 100th and 200th spectroscopic lightcurves from the Eureka! reduction with their best-fit eclipse models, while the bottom two panels show the same for the exoTEDRF reduction. \label{fig:figLC}}
\end{figure*}

\begin{table*}
	\centering
	\caption{Priors of the free parameters used in the atmospheric retrievals. \label{tab:retrieval_priors}}
	\begin{tabular}{l c}
		\hline
		Parameter & Prior \\
		\hline
		$T_{\mathrm{int} [K]}$ & $\mathcal{U}(0,1000)$\\
		$T_{\mathrm{equ} [K]}$ & $\mathcal{U}(0,5000)$\\
		$[\gamma]$ & $\mathcal{U}(-2,2)$\\
		$[\kappa_{\mathrm{IR}}]$ & $\mathcal{U}(-4,0)$\\
		$f_{\mathrm{sed}}$ & $\mathcal{U}(0.1,20)$\\
		$\sigma_{\mathrm{lnorm}}$ & $\mathcal{U}(1,5)$\\
		$[K_{zz}]$ & $\mathcal{U}(2,15)$\\
		$f_c$ & $\mathcal{U}(0,1)$\\
		$[Z/Z_\odot]$ & $\mathcal{U}(-2,2)$\\
		$C/O$ & $\mathcal{U}(0.05,2)$\\
		$[P_q]$ & $\mathcal{U}(-12,2)$\\
		$[\mathrm{H_2O}]$ & $\mathcal{U}(-12,-0.5)$\\
		$[\mathrm{CO}]$ & $\mathcal{U}(-12,-0.2)$\\
		$[\mathrm{CO_2}]$ & $\mathcal{U}(-12,-0.5)$\\
		$[\mathrm{SiO}]$ & $\mathcal{U}(-12,-0.2)$\\
		$[\mathrm{CH_4}]$ & $\mathcal{U}(-12,-1.0)$\\
		$[\mathrm{C_2H_2}]$ & $\mathcal{U}(-12,-1.0)$\\
		$[\mathrm{HCN}]$ & $\mathcal{U}(-12,-1.0)$\\
		$[\mathrm{PH_3}]$ & $\mathcal{U}(-12,-1.0)$\\
		$[\mathrm{H_2S}]$ & $\mathcal{U}(-12,-1.0)$\\
		$[\mathrm{TiO}]$ & $\mathcal{U}(-14,-1.0)$\\
		$[\mathrm{VO}]$ & $\mathcal{U}(-14,-1.0)$\\
		$[\mathrm{MgSiO_3(s)}]$ & $\mathcal{U}(-14,-0.2)$\\
		$[\mathrm{Mg_2SiO_4(s)}]$ & $\mathcal{U}(-14,-0.2)$\\
		$[\mathrm{Al_2O_3(s)}]$ & $\mathcal{U}(-14,-0.2)$\\
		$[\mathrm{SiO_2(s)}]$ & $\mathcal{U}(-14,-0.2)$\\
		$[\mathrm{CaTiO_3(s)}]$ & $\mathcal{U}(-14,-0.2)$\\
		$[\mathrm{TiO_2(s)}]$ & $\mathcal{U}(-14,-0.2)$\\
		\hline
	\end{tabular}
\end{table*}

\begin{figure*}
\includegraphics[width=1\columnwidth]{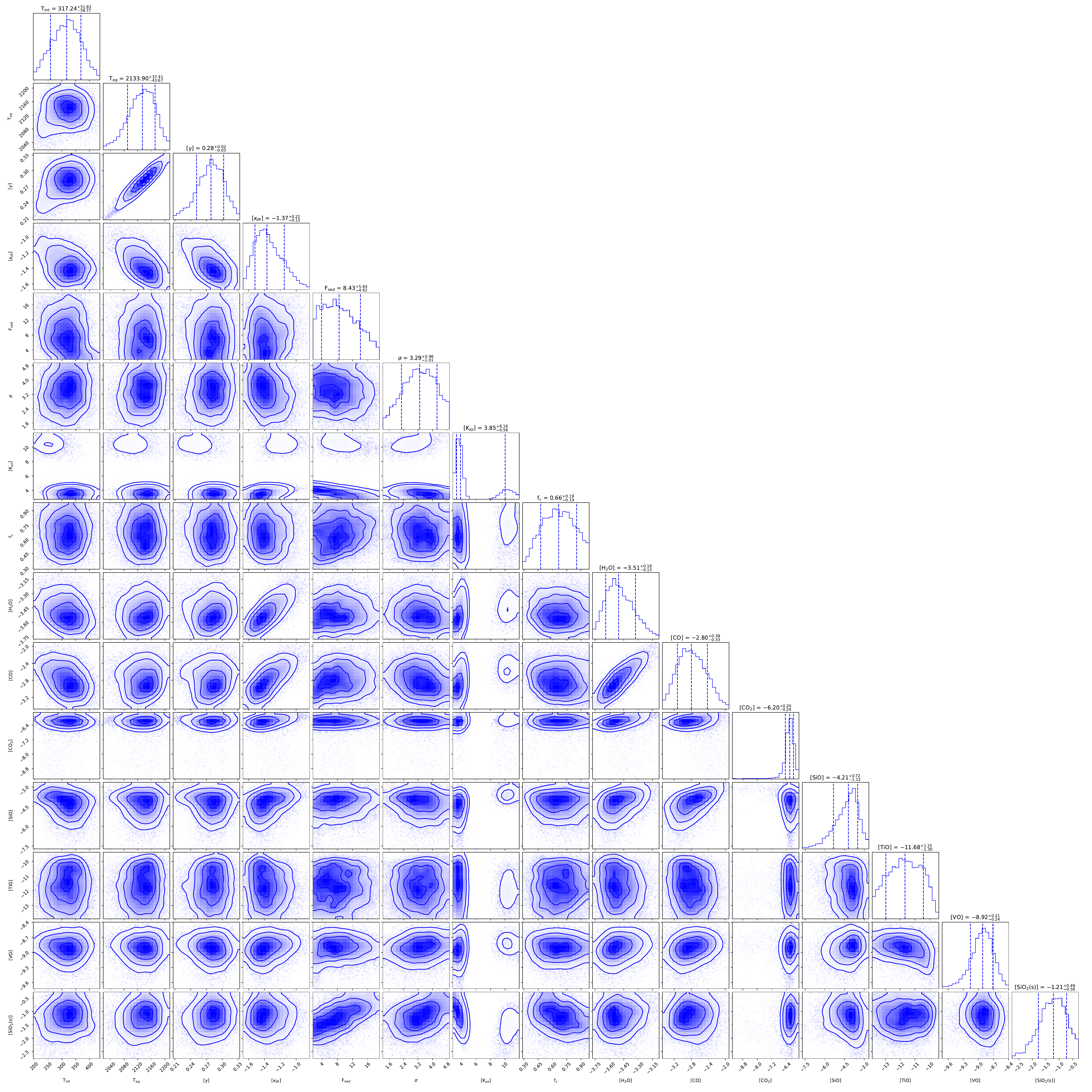}
\caption{Corner plot showing the posterior distributions from the free chemistry retrieval analysis (Eureka!), including only the well-constrained molecular species. \label{fig:figA2}}
\end{figure*}

\begin{figure*}
\includegraphics[width=1\columnwidth]{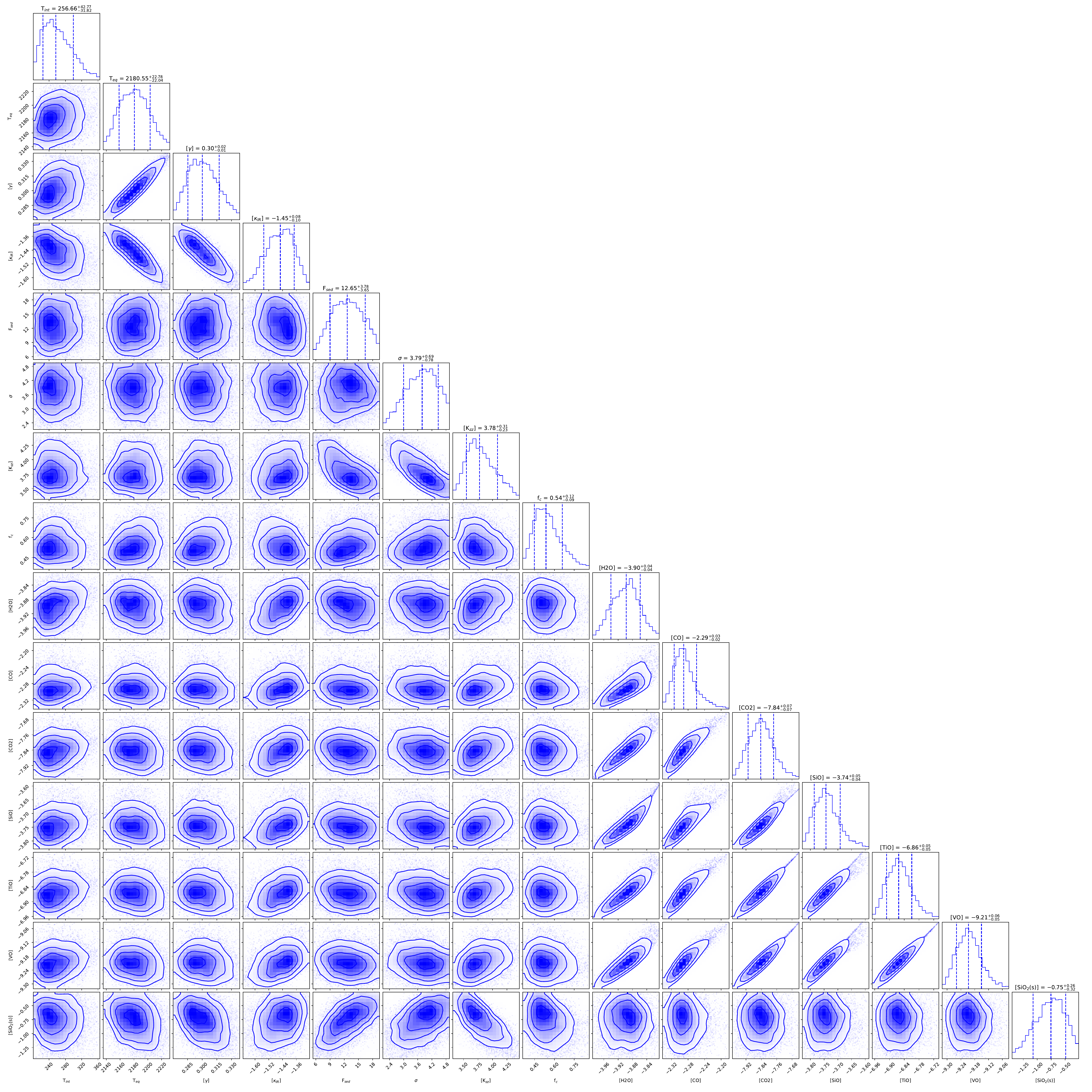}
\caption{Same as Figure \ref{fig:figA2}, but for the equilibrium chemistry retrieval. \label{fig:figA2eq}}
\end{figure*}

\begin{figure*}
\centering
\includegraphics[width=1\columnwidth]{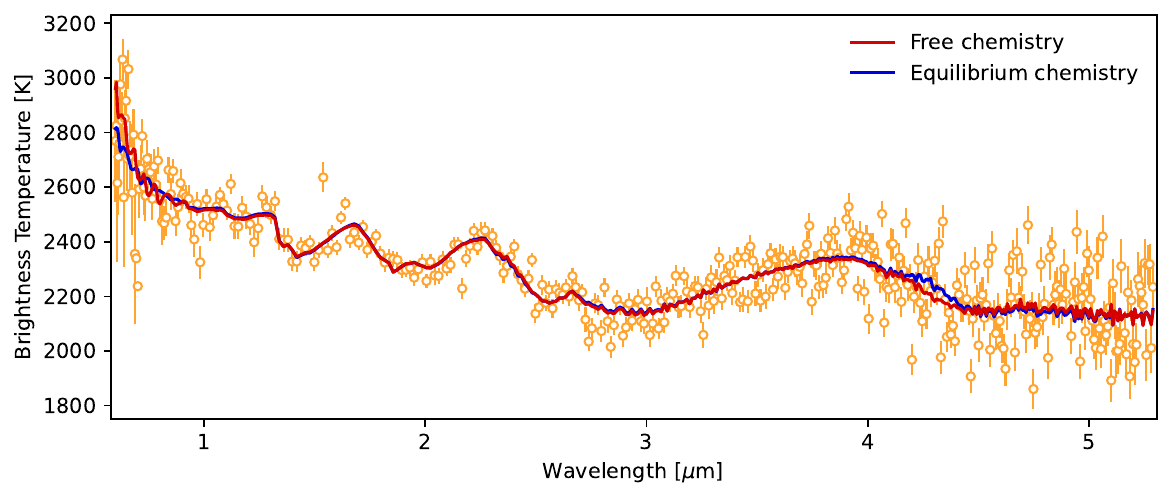}
\caption{The observed emission spectrum of WASP-19b (Eureka!, orange), shown alongside the best-fit free and equilibrium chemistry models, demonstrating strong consistency. \label{fig:figA4}}
\end{figure*}

\begin{figure*}
\centering
\includegraphics[width=1\columnwidth]{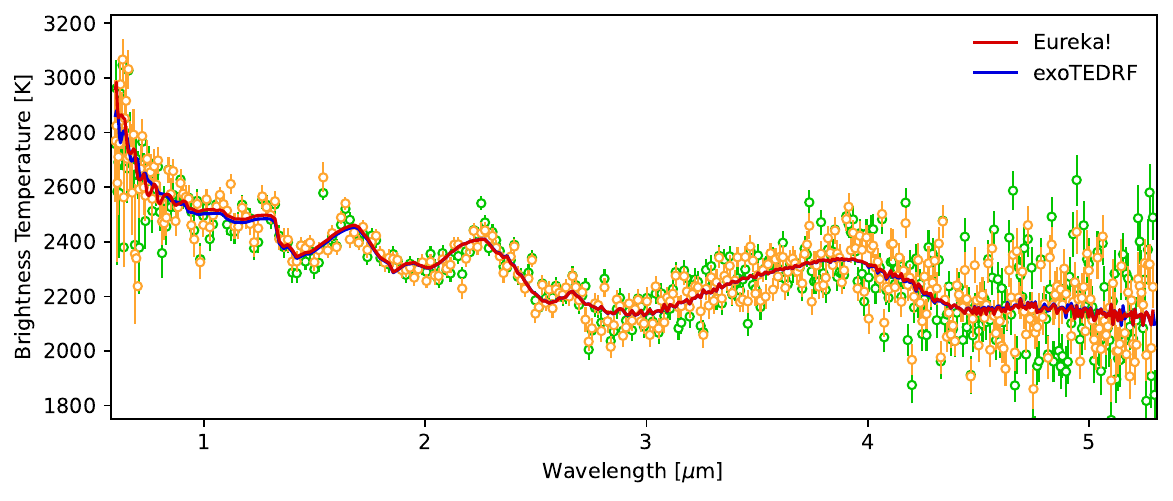}
\caption{The observed emission spectra of WASP-19b from both Eureka! (orange) and exoTEDRF (green) are shown alongside their corresponding best-fit free chemistry models. Both models show strong consistency, validating the robustness of retrieved properties. \label{fig:figA5}}
\end{figure*}

\end{document}